\documentclass[11pt]{article} 
\usepackage{mathrsfs}
\usepackage{amsmath,epsfig,amssymb,epsf,citesort}
\usepackage{url}
\usepackage[left=2cm,top=2cm,bottom=2cm,right=2cm,nohead]{geometry}
\usepackage{amsmath,amssymb,epsf,epsfig,times}
\usepackage[all]{xy}
\usepackage{subfig}
\usepackage{graphicx,color}
\usepackage{url}
\usepackage{cite}
\usepackage{indentfirst}

\usepackage{amsthm}

\DeclareMathOperator*{\argmin}{arg\,min}
\DeclareMathOperator*{\argmax}{arg\,max}

\usepackage{multirow}
\usepackage{dcolumn}
\usepackage{tabularx}

\usepackage{longtable}
\usepackage{lscape}

\usepackage{algorithmicx}
\usepackage[square,sort,comma,numbers]{natbib}
\UseRawInputEncoding

\usepackage{fancyhdr,graphicx,amsmath,amssymb}
\usepackage[linesnumbered,ruled,vlined]{algorithm2e}

\usepackage{algpseudocode}
\usepackage{rotating}
\usepackage{fancybox}
\usepackage{color}

 \usepackage{geometry}
\usepackage{ragged2e}
\usepackage{booktabs, tabularx}
\usepackage{enumitem}
\usepackage{etoolbox}

\usepackage{multirow}

\usepackage{pdflscape}
\usepackage{afterpage}
\usepackage{capt-of}

\usepackage{epigraph}
\usepackage[font=small]{caption}

\usepackage{hyperref}

\definecolor{acronymcolor}{RGB}{232,91,70}

\newtheorem{lemma}{Lemma}

\newtheorem{observation}{Observation}

\theoremstyle{definition}

\newtheorem{remark}{Remark}

\newcommand{\OMIT}[1]{}

\title{\bf{PrEF: \textcolor{acronymcolor}{P}e\textcolor{acronymcolor}{r}colation-based \textcolor{acronymcolor}{E}volutionary \textcolor{acronymcolor}{F}ramework for the diffusion-source-localization problem in large networks}}
\author{Yang Liu$^{1}$\footnote{yangliuyh@gmail.com}\ , Xiaoqi Wang$^1$, Xi Wang$^{2,3}$, Zhen Wang$^1$, J$\ddot{\text{u}}$rgen Kurths$^4$\\
$^1${\emph{Northwestern Polytechnical University}}\\
$^2${\emph{The Chinese University of Hong Kong}}\\
$^3${\emph{Stanford University}}\\
$^4${\emph{Potsdam Institute for Climate Impact Research}}
}
\date{}

\begin{document}


\maketitle

\begin{abstract}

    We assume that the state of a number of nodes in a network could be investigated if necessary, and study what configuration of those nodes could facilitate a better solution for the diffusion-source-localization (DSL) problem. In particular, we formulate a candidate set which contains the diffusion source for sure, and propose the method, \textbf{P}e\textbf{r}colation-based \textbf{E}volutionary \textbf{F}ramework (\textbf{PrEF}), to minimize such set. Hence one could further conduct more intensive investigation on only a few nodes to target the source. To achieve that, we first demonstrate that there are some similarities between the DSL problem and the network immunization problem. We find that the minimization of the candidate set is equivalent to the minimization of the order parameter if we view the observer set as the removal node set. Hence, PrEF is developed based on the network percolation and evolutionary algorithm. The effectiveness of the proposed method is validated on both model and empirical networks in regard to varied circumstances. Our results show that the developed approach could achieve a much smaller candidate set compared to the state of the art in almost all cases. Meanwhile, our approach is also more stable, i.e., it has similar performance irrespective of varied infection probabilities, diffusion models, and outbreak ranges. More importantly, our approach might provide a new framework to tackle the DSL problem in extreme large networks.

\end{abstract}

%

\date{}


\vspace{0.2cm}

\section{Introduction}

There has recently been an enormous amount of interest focusing on the diffusion-source-localization (DSL) problem on networks, which aims to find out the real source of an undergoing or finished diffusion \cite{shah2010detecting, pinto2012locating, ali2019epa}. Two specific scenarios are epidemics and misinformation and both of which can be well modeled by the networks. Being one of the biggest enemies to global health, infectious diseases could cause rapid population declines or species extinction \cite{harvell2002climate}, from the Black Death (probably bubonic plague), which is estimated to have caused the death of as much as one third of the population of Europe between 1346 and 1350 \cite{brauer2012mathematical}, to nowadays COVID-19 pandemic, which might result in the largest global recession in history, in particular, climate change keeping exacerbating the spread of diseases and increasing the probability of global epidemics \cite{mcmichael2003climate, jamison2013global}. In this case, the study of the DSL problem can potentially help administrations to make policies to prevent future outbreaks and hence save a lot of lives and resources. Further regarding misinformation, as one of the biggest threats to our societies, it could cause great impact on global health and economy and further weaken the public trust in governments \cite{zhou2020survey}, such as the ongoing COVID-19 where fake news circulates on online social networks (OSNs) and has made people believe that the virus could be killed by drinking slaty water and bleach \cite{sahoo2020demystifying}, and the `Barack Obama was injured' which wiped out \$130 billion in stock value \cite{fakenewseconomy}. In this circumstance, the localization of the real source would play important role for containing the misinformation fundamentally.

The main task of the DSL problem is to find an estimator that could give us an inferred source based on the known partial information, and the most ideal estimator is the one that gives us the real source. However, due to the complexity of contact patterns and the uncertainty of diffusion models, the real source is generally almost impossible to be inferred exactly, even the underlying network is a tree \cite{choi2020information}. Hence, as an alternative, the error distance is used, and an estimator is said to be better than another one if the corresponding inferred source is closer to the real source in hop-distance \cite{pinto2012locating, zhu2014information, wang2014rumor, jiang2016identifying, ali2019epa, choi2020information}. And therefore, varied methods have been developed to minimize such error distance based on different-known-information assumptions, such as observers having knowledge of time stamps and infection directions \cite{pinto2012locating}, the diffusion information \cite{lokhov2014inferring}, and the states of all nodes \cite{ali2019epa}, etc. \cite{jiang2016identifying, choi2020information}. But here we argue that: what should we do next once we acquire the estimator having small error distance? Indeed, one can carry out intensive detection on the neighbor region of the inferred source to search for the real source. In this case, for regular networks, a small error distance is usually associated with a small number of nodes to be checked. However, most real-world networks are heterogeneous, which indicates that even a short error distance might correspond to a great number of nodes, especially those social networks.

Hence, in this paper, we present a method, Percolation-based Evolutionary Framework (PrEF), to tackle the DSL problem by suppressing a candidate set that the real source belongs to for sure. In particular, we assume that there are a group of nodes in the networks, whose states can be investigated if necessary. Meanwhile, those nodes are also assumed to have information of both time stamps and infection directions. Then, our goal is to use as fewer observers (nodes) as possible to achieve the containment of the candidate set. We find that such goal can be reached by the solution of the network immunization problem. Hence, we have our method based on the network percolation and evolutionary computation. Results on both model and empirical networks show that the proposed method is much better compared to the state-of-the-art approaches.


Key contributions of this paper are summarized as follows:
\begin{itemize}
  \item \textbf{DSL vs. network immunization.} We concretely study and derive the connection of the DSL problem and the network immunization problem, and find that the solution of the network immunization problem can be used to and could effectively cope with the DSL problem.
  \item \textbf{Percolation-based evolutionary framework.} We propose a percolation-based evolutionary framework to solve the DSL problem, which takes a network percolation process and potentially works for a range of scenarios.
  \item \textbf{Extensive evaluation on synthetic and empirical networks.} We evaluate the proposed method on two synthetic networks and four empirical networks drawn from different real-world scenarios, whose sizes are up to over $800,000$ nodes. Results show that our method is more effective, efficient, and stable than the-state-of-the-art approaches, and is also capable of handling large networks.
\end{itemize}

\section{Related Work}

\textbf{DSL approaches.} Shah and Zaman first studied the DSL problem of single diffusion source and introduced the rumor centrality method by counting the distinct diffusion configurations of each node \cite{shah2010detecting, shah2012rumor}. They considered the Susceptible-Infected (SI) model and concluded that a node is said to be the source with higher probability if it has a larger rumor centrality. Following that, Dong et al. also investigated the SI model and proposed a better approach based on the local rumor center generalized from the rumor centrality, given that a priori distribution is known for each node being the rumor source \cite{dong2013rooting}. Similarly, Zhu and Ying investigated the problem under the Susceptible-Infected-Recovered (SIR) model and found that the Jordan center could be used to characterize such probability \cite{zhu2014information}. Wang et al. showed that the detection probability could be boosted if multiple diffusion snapshots were observed \cite{wang2014rumor}, which can be viewed as the case that the information of the diffusion time was integrated to some extent. Indeed, if the time-stamp or other additional information is known, the corresponding method would usually work better \cite{pinto2012locating, ali2019epa, chai2021information}. In short, almost all those exiting methods study the DSL problem on either simple networks (such a tree-type networks or model networks) or small empirical networks. Hence, their performance might be questioned in real and complex scenarios, such as networks having a lot of cycles \cite{watts1998collective, newman2018networks}.

\textbf{Network immunization.} The network immunization problem aims to find a key group of nodes whose removal could fragment a given network to the greatest extend, which has been proved to be a NP-hard problem \cite{morone2015influence}. In general, approaches for coping with this problem can be summarized into four categories. The first one is to obtain the key group by strategies such as randomly selecting nodes from the network, which is usually called local-information-based methods \cite{cohen2003efficient, liu2016local}. In this scope, since the network topology information does not have to be precisely known, these methods would be quite useful in some scenarios. Rather than that, when the network topology is known, the second category is usually much more effective. Methods of the second category draw the key group by directly classifying nodes based on measurements like degree centrality, eigenvector centrality, pagerank, and betweenness centrality \cite{newman2018networks}. More concretely, they firstly calculate the importance of each node using their centralities and choose those ranking on the top as the key group. The third category takes the same strategy but will heuristically update the importance of nodes in the remaining network after the most important node is removed, and the key group eventually consists of all removed nodes \cite{morone2015influence}. The last category obtains the key group in indirect ways \cite{ren2019generalized, fan2020finding}. For instance, refs. \cite{mugisha2016identifying, braunstein2016network} achieve the goal by tackling the feedback vertex set problem.

\section{Model}

We assume that a diffusion $\zeta$ occurs on a network $G(\mathcal{V}, \mathcal{E})$, where $\mathcal{V}$ and $\mathcal{E}$ are the corresponding node set and edge set, respectively. Letting $e_{uv} \in \mathcal{E}$ be the edge between nodes $u$ and $v$, we define the nearest neighbor set regarding $u$ as $\Gamma(u)=\{v, \forall e_{uv} \in \mathcal{E}\}$. A connected component $c_i$ of $G$ is a subnetwork $G'(\mathcal{V}', \mathcal{E}')$ satisfying $\mathcal{V}' \subset \mathcal{V}$, $\mathcal{E}' = \mathcal{E} \cap (\mathcal{V}' \times \mathcal{V}')$, and $\mathcal{E} \cap ( (\mathcal{V}\setminus\mathcal{V}') \times \mathcal{V}')\equiv\emptyset$. In particular, denoting $|c_i|$ the size of $c_i$ (i.e., the number of nodes in $c_i$), the largest connected component (LCC), $c_{\text{max}}$, is then defined as the component that consists of the majority of nodes, that is, $c_{\text{max}} = \argmax_{c_i} |c_i|$. Now assuming that $G''$ is the remaining network of $G$ after the removal of $q$ fraction nodes and the incident edges, the corresponding size of the LCC, $|c''_{\text{max}}|$, will hence be a monotonically decreasing function of $q$. Such function is also known as the order parameter $\mathcal{G}(q)=|c''_{\text{max}}|/n$, where $n=|\mathcal{V}|$ is the number of nodes in $G$. According to the percolation theory \cite{stauffer2018introduction}, when $q$ is large enough, say larger than the critical threshold $q_c$, the probability that a randomly selected node belongs to the LCC would approach zero. In other words, if $q>q_c$, then there would be no any giant connected component in $G''$.


The diffusion $\zeta$ is generally associated with four factors: the network structure $G$, the diffusion source $v_s$, the dynamic model $M$, and the corresponding time $t$, say $\zeta(G, v_s, M, t)$. Regarding $M$, here we particularly consider the Susceptible-Infected-Recovered (SIR) model \cite{keeling2011modeling} as an example to explain the proposed method. More models will further be discussed in the result section. For nodes of $G$ governed by the SIR model, their states are either susceptible, infected or recovered. As $t\rightarrow t+1$, an infected node $u$ has an infection probability $\beta_{uv}$ (or a time interval $\tau_{uv}=1/\beta_{uv}$) to transmit the information or virus, say $\varsigma$, to its susceptible neighbor $v \in \Gamma(u)$. Meanwhile, it recovers with a recovery probability $\gamma_u$ (or duration of $\tau_u=1/\gamma_u$). Those recovered nodes stay in the recovered state and $\varsigma$ cannot pass through a recovered node.

Now supposing that a group of nodes $\mathcal{O} \in \mathcal{V}$ are particularly chosen as observers and hence their states could be investigated if necessary, we study what and how the configuration of $\mathcal{O}$ could facilitate better solutions for the diffusion-source-localization (DSL) problem \cite{shah2010detecting, pinto2012locating, ali2019epa}. In particular, we assume that a node $u \in \mathcal{O}$ would record the relative infection time stamp $t_u$ once it gets infected. Besides, we also consider that part of $\mathcal{O}$, say $\mathcal{O}_d$, have the ability to record the infection direction, that is, a node $u \in \mathcal{O}_d$ can also show us the node $v$ if $v$ transmits $\varsigma$ to $u$. Based on these assumptions, for a diffusion $\zeta$ triggered by an unknown diffusion source $u_s$ at time $t_0$, the DSL problem aims to design an estimator $\sigma(G, \mathcal{O})$ that gives us the inferred source $\widehat{u}_s = \sigma(G, \mathcal{O})$ satisfying $\widehat{u}_s =\argmax_{v \in G} \mathcal{P}(\mathcal{O}|v)$, where $\mathcal{P}(\mathcal{O}|v)$ is the probability that we observe $\mathcal{O}$ given $\zeta(G, v, M, t)$. Obviously, the state of a node $i\in \mathcal{O}$ is governed by all parameters of $\zeta$ but with unknown $M$ and $t$. Hence, with high probability $\widehat{v}_s$ differs from the real source $v_s$ in most scenarios \cite{pinto2012locating, zhu2014information, ali2019epa}. And therefore, the error distance $\epsilon=h(\widehat{v}_s, v_s)$ is conducted to verify the performance of an estimator, where $h(\widehat{v}_s, v_s)$ represents the hop-distance between $\widehat{v}_s$ and $v_s$. Usually, an estimator is said to be better than another one if it has smaller $\epsilon$.

But here we argue that: what should we do next after we obtain the estimator having a small $\epsilon$? Or in other words, can the estimator help us find the diffusion source more easily? Indeed, after acquiring $\widehat{v}_s$, one can further conduct more intensive search on the neighbor region of $\widehat{v}_s$ to detect $v_s$. In this case, a small $\epsilon$ generally corresponds to a small search region. However, due to the heterogeneity of contact patterns in most real-world networks \cite{barabasi1999emergence}, even a small region (i.e., a small $\epsilon$) might be associated with a lot of nodes. Therefore, it would be more practical in real scenarios if such estimator gives us a candidate set $\mathcal{V}_c$ satisfying $v_s \in \mathcal{V}_c$ for sure. And hence, we formulate the goal function that this paper aims to achieve as
\begin{equation}\label{eq_model_1}
  \widehat{\mathcal{O}}=\argmin_{\mathcal{O}} |\mathcal{V}_c|, v_s \in \mathcal{V}_c \text{ for sure},
\end{equation}
where $|\mathcal{V}_c|$ is the size of $\mathcal{V}_c$. Intuitively, Eq. (\ref{eq_model_1}) is developed based on the assumption that: the smaller the candidate set, the lower the cost of the intensive search. And in general, $\mathcal{V}_c$ should be finite guaranteed by finite $\mathcal{O}$ for an infinite $G$, otherwise, the cost would be infinite since the intensive search has to be carried out on an infinite population.



%
%




\section{Method}

Let $\mathcal{V}_r$ be the removal node set and $\mathcal{V}_o$ the rest, i.e., $\mathcal{V}_r \cup \mathcal{V}_o = \mathcal{V}$ and $\mathcal{V}_r \cap \mathcal{V}_o = \emptyset$. For the subnetwork $G'$ regarding $\mathcal{V}_o$, the boundary of a connected component $c_i$, say $\widehat{c}_i$, is defined as
\begin{equation}\label{eq_method_1}
  \widehat{c}_i = \{u | e_{uv} \in \mathcal{E}, \forall v\in c_i, \forall u\in \mathcal{V}_r\}.
\end{equation}
Likewise, we write the component cover $\alpha(u)$ that a specific node $u\in \mathcal{V}_r$ corresponds to as
\begin{equation}\label{eq_method_2}
  \alpha(u) = \bigcup_{v\in \Gamma(u)} c_i(v),
\end{equation}
where $c_i(v)$ represents the component that node $v$ belongs to. Denoting $t_v$ the time stamp that node $v$ gets infected and $\mathcal{O}' = \{u| u=\argmin_{v\in \mathcal{O}} t_v \}$, where $t_v$ is assumed to be infinite if $v$ is still susceptible, the proposed approach is developed based on the following observation (Observation \ref{obser1}).
\begin{observation}\label{obser1}
    Letting
    \begin{equation}\label{eq_method_3}
        \mathcal{V}_c' =  \bigcap_{\forall u \in \mathcal{O}'} \alpha(u) \cup \{u\},
    \end{equation}
    we then have $v_s \in \mathcal{V}_c'$ for sure.
\end{observation}


\begin{figure}[htb]
  \centering
  \includegraphics[width=0.95\linewidth]{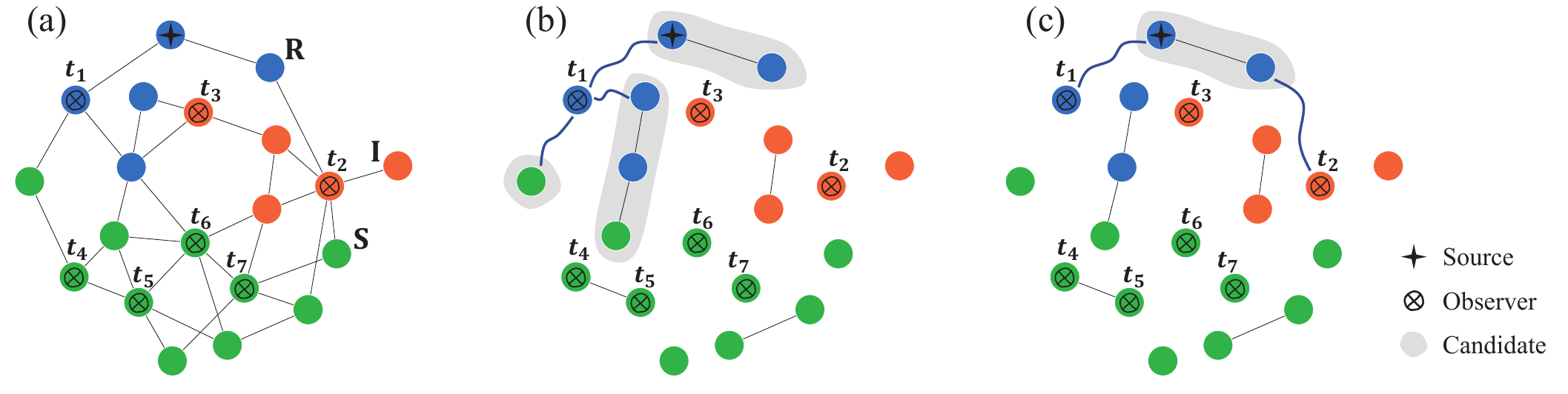}
  \caption[]{Examples regarding DSL, where $\mathcal{V}_r=\mathcal{O}$ (Observer) and $\mathcal{V}_o$ the rest. Nodes in susceptible state are colored by green (marked by $\mathbf{S}$), infected by orange ($\mathbf{I}$), and recovered by blue ($\mathbf{R}$). $t_v$ represents the time stamp that node $v$ gets infected, e.g., $t_1$ of node $1$. (a) Snapshot of $\zeta$. (b) $\mathcal{O}'=\{1\}$, i.e., $t_1<t_v, v\in [2, 7]$. (c) $\mathcal{O}'=\{1, 2\}$, i.e., $t_1 = t_2 < t_v, v\in [3, 7]$.}
  \label{fig_method_example_1}
\end{figure}

Example 1. Considering Fig. \ref{fig_method_example_1}(a) and (b), the boundary of the connected component that the diffusion source belongs to is $\{1, 2\}$.

Example 2. Regarding Fig. \ref{fig_method_example_1}(b), $\alpha(1)$ consists of all nodes covered by those shadows.

Example 3. With respect to Fig. \ref{fig_method_example_1}(c), $\mathcal{V}_c'$ comprises of node $1$, node $2$, the diffusion source, and the node adjacent to the source. {\hfill $\blacksquare$\par}

We now consider the generation of the \textbf{observer set $\mathcal{O}$ (or equivalently $\mathcal{V}_r$)}. For a given network $G(\mathcal{V}, \mathcal{E})$ constructed by the configuration model \cite{molloy1995critical}, letting $\langle k^2\rangle$ and $\langle k\rangle$ accordingly be the first and second moments of the corresponding degree sequence, we then have Lemma \ref{method_lemma_1}.
\begin{lemma}\label{method_lemma_1}
  (Molloy-Reed criterion \cite{molloy1995critical}) A network $G$ constructed based on the configuration model with high probability has a giant connected component (GCC) if
  \begin{equation}\label{eq_method_4}
    \langle k^2\rangle / \langle k\rangle > 2,
  \end{equation}
  where GCC represents a connected component whose size is proportional to the network size $n$.
\end{lemma}
Now suppose that $\mathcal{V}_r$ consists of nodes randomly chosen from $G$ and let $q=|\mathcal{V}_r|$ represent the fraction of removed nodes. Apparently, such removal would change the degree sequence of the remaining network (i.e., subnetwork $G'$, also see Fig. \ref{fig_method_example_1}(b) as an example). Assuming that there is a node $v$ shared by both $G$ and $G'$, the probability that its degree $k$ (in $G$) decreases to a specific degree $k'$ (in $G'$) should be
$$\tbinom{k}{k'} (1-q)^{k'}q^{k-k'},$$
where $\tbinom{k}{k'}$ is the combinational factor (note that each node has $q$ probability of being removed). Letting $p_k$ denote the degree distribution of $G$, we then have the new degree distribution $p'_{k'}$ of $G'$ as
\begin{equation}\label{eq_method_5}
p'_{k'}=\sum_{k=k'}^\infty p_k \tbinom{k}{k'} (1-q)^{k'}q^{k-k'},
\end{equation}
and hence the corresponding first and second moments can be further obtained as
\begin{equation}\label{eq_method_6}
\langle k' \rangle=\sum_{k'=0}^\infty k' p'_{k'} =\sum_{k'=0}^\infty k' \sum_{k=k'}^\infty p_k \tbinom{k}{k'} (1-q)^{k'}q^{k-k'}=(1-q)\langle k \rangle
\end{equation}
and
\begin{equation}\label{eq_percolation_rob_7}
\langle k'^2 \rangle=(1-q)^2\langle k^2 \rangle + q(1-q)\langle k \rangle,
\end{equation}
respectively. Since $G$ is constructed using the configuration model, its edges are independent of each other. That is, each edge of $G$ shares the same probability of connecting to $v \in \mathcal{V}_r$. In other words, the removal of $v$ would remove each edge with the same probability and hence $G'$ can be viewed as a special network that is also constructed by the configuration model. Thus, Lemma \ref{method_lemma_1} can be used to determine whether a GCC exists in $G'$ and we reach
\begin{equation}\label{eq_percolation_rob_8}
q_c=1-\frac{1}{\langle k^2 \rangle/\langle k \rangle -1},
\end{equation}
where $q_c$ is the critical threshold of $q$, that is: i) if $q<q_c$, with high probability there is a GCC in $G'$; ii) if $q>q_c$, with high probability there is no GCC in $G'$ \cite{cohen2000resilience, barabasi2016network}.

For random networks (such as Erd{\H{o}}s-R{\'e}nyi (ER) networks \cite{erdds1959random}), $\langle k^2 \rangle = \langle k \rangle(\langle k \rangle + 1)$ gives us $q_c=1-1/\langle k \rangle$, which indicates that $q_c$ is usually less than $1$ and it increases as $G$ becomes denser. But for heterogeneous networks, say $p_k\backsim k^{-\ell}$, $\langle k^2 \rangle=\sum_k^\infty k^2p_k \backsim \sum_k k^{2-\ell}$ diverges if $\ell<3$ (most empirical networks have $2<\ell<3$ \cite{newman2018networks}), which means $q_c$ approaches $1$.

\begin{remark}
From the above analysis, we have the following conclusions regarding the case that the observer set $\mathcal{O}$ is randomly chosen. For random networks, $q_c<1$ indicates that one can always have a proper $\mathcal{O}$ to achieve finite $\mathcal{V}_c'$. And usually the denser the network, the larger the observer set $\mathcal{O}$. But for heterogeneous networks, $q_c \rightarrow 1$ means that such goal could only be achieved by putting almost all nodes into the observer set $\mathcal{O}$. {\hfill $\blacksquare$ \par}
\end{remark}

We further consider the case that $\mathcal{O}$ consists of hubs \cite{albert2000error}, where hubs represent those nodes that have more connections in a particular network. Specifically, for a network $G(\mathcal{V}, \mathcal{E})$, we first define a sequence $\mathcal{S}$ regarding $\mathcal{V}$ and assume that each element of $\mathcal{S}$ is uniquely associated with a node in $G$. Then, letting $\mathcal{S}(i)=u$ and $\mathcal{S}(j)=v$ if $k_u\geqslant k_v$, satisfying $i<j$, where $k_u$ represents the degree of node $u$, we have $\mathcal{O}=\{\mathcal{S}(i), i\in [1, \lfloor nq+0.5\rfloor]\}$. For heterogeneous networks under such removals on hubs, threshold $q_c<1$ can be achieved and obtained by numerically solving \cite{cohen2001breakdown}
\begin{equation}\label{eq_percolation_rob_9}
q_c^{(2-\ell)/(1-\ell)} - 2= \frac{2-\ell}{3-\ell}k_{\text{min}}(q_c^{(3-\ell)/(1-\ell)}-1),
\end{equation}
where $k_{\text{min}}$ is the minimum degree.

For $\mathcal{V}_c'$, however, we could not obtain an explicit equation to indicate whether it is finite. But we can roughly show that there should be $q_c<1$ that gives us a finite $\mathcal{V}_c'$ for networks generated by the configuration model with degree distribution $p_k\backsim k^{-\ell}$. Supposing that the size of the LCC of $G'$ is proportional to $n^b$ with $b<1$, then
\begin{equation}\label{eq_percolation_rob_900}
k_{\text{max}}n^b/n
\end{equation}
gives us the possibly largest size of $\mathcal{V}_c'$, where $k_{\text{max}}$ is the maximum degree of $G$ and here we assume that such degree is unique. In the mentioned case, $k_{\text{max}}=k_{\text{min}} n^{1/(\ell-1)}$ holds and hence it approaches $0$ as $n\rightarrow \infty$ if $2<\ell<3$ (again, this is the case that we are particularly interested in), which indicates that one can always find some proper value of $b>0$ satisfying $b<(\ell-2)/(\ell-1)$ for a given $\ell$. Note that, same as the random removal, each edge of $G$ also shares the same probability of being removed. Hence, both the size of the LCC and the number of connections that node $v_{\text{max}}$ has with $\mathcal{V}_o$ decrease as $q$ increases, where $v_{\text{max}}$ represents the node whose degree is $k_{\text{max}}$.

But for networks having $k_{\text{max}} \backsim n$ such as star-shape networks, $\mathcal{V}_c'$ is still infinite when $q$ is finite. In this case, the information of only infection time stamps of nodes in $\mathcal{O}$ is apparently not enough. However, if the infection direction is also recorded, then the central node as the unique observer is fairly enough for the DSL problem in those star-shape networks. Note that most existing DSL methods would also fail in star-shape networks. Hence, we further assume that part of $\mathcal{O}$, say $\mathcal{O}_d$, can also show the infection direction. Obviously, $|\mathcal{V}_c'|$ is a monotonically decreasing function of $|\mathcal{O}_d|$ for a specific $q$. In particular, if $\mathcal{O}_d=\mathcal{O}$, then the size of $\mathcal{V}_c'$ would be bounded by the size of the LCC.

\begin{remark}
In general, finite $\mathcal{V}_c'$ of heterogeneous networks can be achieved by carefully choosing $\mathcal{O}$. Or in other words, the configuration of $\mathcal{O}$ plays fundamental role for the suppression of $\mathcal{V}_c'$. Associating $\mathcal{O}$ with the sequence $\mathcal{S}$, such as random removal can be viewed as a removal over a random sequence $\mathcal{S}$, our goal is now to acquire a better $\mathcal{S}$ that could give us a smaller $\mathcal{V}_c'$ in regard to a specific $q$. Besides, since the number of components that $v_{\text{max}}$ connects to is usually difficult to measure especially for real-world networks, here we choose to achieve the containment of $\mathcal{V}_c'$ by curbing the LCC of $G'$ (see also Eq. (\ref{eq_percolation_rob_900})), which coincides with the suppression of the order parameter $\mathcal{G}(q)$. {\hfill $\blacksquare$\par}
\end{remark}

Therefore, we reach
\begin{equation}\label{eq_percolation_rob_10}
  \mathcal{\widehat{S}}=\left\{
\begin{aligned}
& \argmin_\mathcal{S} q_c(\mathcal{S}), \ \ \text{if $\delta$ is given}, \\
& \argmin_\mathcal{S} F(\mathcal{S}), \ \ \text{otherwise},
\end{aligned}
\right.
\end{equation}
where $q_c(\mathcal{S})=\argmin_q \mathcal{G}(q)\leqslant\delta$, $F(\mathcal{S}) = \sum_q \mathcal{G}(q)$, and $\delta$ is a given parameter (such as $\delta=0.01$). Eq. (\ref{eq_percolation_rob_10}) is also known as the network immunization problem which aims to contain epidemics by the isolation of as fewer nodes as possible \cite{barabasi2016network}. And a lot of methods have been proposed to cope with such problem \cite{morone2015influence, braunstein2016network, clusella2016immunization, liu2018optimization, liu2019framework, fan2019dismantle, ren2019generalized, fan2020finding}. Here we particularly choose and consider the approach based on the evolutionary framework (AEF) to construct $\mathcal{S}$ since it can achieve the state of the art in most networks.


We first introduce several auxiliary variables to the ease of description of AEF. Consider a given network $G(\mathcal{V}, \mathcal{E})$ and the corresponding sequence $\mathcal{S}$. Let $\mathcal{S}_i^\bot=(\mathcal{S}_i, \mathcal{S}_{i+1}, ..., \mathcal{S}_h)$ , where $\mathcal{S}_i$ is a subsequence of $\mathcal{S}$. Likewise, denote $G'(\mathcal{S}_i^\bot)$ a subnetwork $G'(\mathcal{V}', \mathcal{E}')$, in which $\mathcal{V}'=\{\mathcal{S}_i^\bot(j), \forall j\}$ and $\mathcal{E}' = \mathcal{E} \cap (\mathcal{V}' \times \mathcal{V}')$. Based on that, $F$ of $G'(\mathcal{S}_i^\bot)$ regarding $\mathcal{S}_i^\bot$ is written as $F'(\mathcal{S}_i^\bot)$. Further, letting $p'_\text{max}=|\text{LCC}|/n$ of $G'$, we define the critical subsequence $\mathcal{S}_c$ as the subsequence satisfying that $p'_\text{max}\leqslant \delta$ of $G'(\mathcal{S}_i^\bot)$ and $p'_\text{max}> \delta$ of $G'((\mathcal{S}_c,\mathcal{S}_i^\bot))$. Note that all $F'$ is scaled by $n$, namely, the size of the studied network $G$.




The core of AEF is the relationship-related (RR) strategy that works by repeatedly pruning the whole sequence $\mathcal{S}$. Specifically, per iteration $T$, RR keeps a new sequence $\mathcal{S}'$ (i.e., $\mathcal{S}\leftarrow \mathcal{S}'$) if $F(\mathcal{S}')<F(\mathcal{S})$ (or $q_c(\mathcal{S}')<q_c(\mathcal{S})$), which is obtained through the following steps. 1) Let $j=n$, $\mathcal{S}'\leftarrow \mathcal{S}$, and $G'(\mathcal{V}', \mathcal{E}')$ be a subnetwork of $G$, which consists of all nodes in $\mathcal{V}'=\{\mathcal{S}'(z), z\in [j,n]\}$ and the associated edges in $\mathcal{E}'=\{e_{uv}, \forall u, v \in \mathcal{V}'\}$. 2) Construct the candidate set $\bar{s}_j$ by randomly choosing $\Delta$ times from $\{S'(i), i\in [\max(j-r\times n, 1), j]\}$, where $\Delta\in [1, \widehat{\Delta}]$ and $r\in (0, \widehat{r}]$ are randomly generated per iteration, and $\widehat{\Delta}$ and $\widehat{r}$ are given parameters. 3) Choose the node $u$,
\begin{equation}\label{eq_percolation_rob_11}
u = \argmin_v \xi(v), \forall v \in \bar{s}_i,
\end{equation}
where $\xi(v)=\sum_{c_i \in c(v)} |c_i|$ or $\xi(v)=\prod_{c_i \in c(v)} |c_i|$, in which $c(v)$ is the component set that node $v$ would connect. 4) Update $G'$ and $\mathcal{S}'$, namely, $\mathcal{V}' \leftarrow \mathcal{V}' \cup \{u\}$, $\mathcal{E}' \leftarrow \mathcal{E}' \cup \{e_{uv}, \forall v \in \mathcal{V}', v\neq u\}$, and swap $\mathcal{S}'(j)$ and $\mathcal{S}'(z)$ satisfying $\mathcal{S}'(z)=u$. 5) $j\leftarrow j-1$. 6) Repeat steps 2) - 5) until $j=1$, which accounts for one round (see also Algorithm \ref{Alg_framework_basic_method_1}). And RR acquires the solution by repeating steps 1) - 6) $\widehat{T}$ times.

\noindent\makebox[\textwidth][c]{
	\small
	\begin{minipage}[c]{.7\linewidth}
		\vspace{\textfloatsep}
		\begin{algorithm}[H]
			\caption{One round of RR \cite{liu2018optimization}}\label{Alg_framework_basic_method_1}
			\KwIn{$G(\mathcal{V}, \mathcal{E})$, $\mathcal{S}$, $\widehat{\Delta}$, $\widehat{r}$}
			\KwOut{$\mathcal{S}$}
			\textbf{Initialization:} $\mathcal{V} \leftarrow \{\}$, $\mathcal{E} \leftarrow \{\}$, $j \leftarrow n, \mathcal{S}' \leftarrow \mathcal{S}$, $\Delta$, and $r$\\
			\While{$j \geqslant 1$}{
				$j \leftarrow j-1$\\
				Get the candidate set $\bar{s}_i$ based on $\Delta$ and $r$\\
				Choose node $u \in \bar{s}_i$ based on Eq. (\ref{eq_percolation_rob_11})\\
				Update $G'(\mathcal{V}', \mathcal{E}')$ and $\mathcal{S}'$\\
			}
			\If{$F(\mathcal{S}')<F(\mathcal{S})$}{
				$\mathcal{S} \leftarrow \mathcal{\mathcal{S}}'$
			}
		\end{algorithm}
		\vspace{\textfloatsep}
\end{minipage}}

\begin{observation}\label{obser2}
	Supposing that\begin{equation}\label{eq_percolation_rob_12}
    	F_i'=F'(\mathcal{S}_i^\bot \leftarrow \mathcal{S}_{i+1}^\bot)= F'(\mathcal{S}_i^\bot) - F'(\mathcal{S}_{i+1}^\bot)
    \end{equation}
	holds, then for a specific sequence $\mathcal{S}$ regarding a given network $G$, $F_i'$ would be independent of $F_j'$ if either $i>j$ or $j>i$ satisfies. That is, for such a case, the order of nodes in $\mathcal{S}_i$ has no effect on $F_j'$. {\hfill $\blacksquare$\par}
\end{observation}

AEF is developed based on RR and Observation \ref{obser2}. Specifically, at $T_p$, a random integer $j(T_p)\in [\pi_1, \pi_2]$ is generated, where $\pi_1$ and $\pi_2$ are two given boundaries. Let $\mathcal{S}_i=(\mathcal{S}(z)), \forall z \in [j(T_p)\times (i-1) + 1, \min(j(T_p)\times i)]$. Then, for all subsequences $\mathcal{S}_i, \forall i \in [1, h]$, RR with the optimization of $F$ is conducted if $\delta$ is unknown, otherwise, $\mathcal{S}_c$ is optimized by RR with $q_c$ minimum.



Hence, the containment of $\mathcal{V}'$ has been achieved (see also Eq. (\ref{eq_method_3})), based on which existing DSL approaches can be used to further acquire the candidate set $\mathcal{V}_c \subset \mathcal{V}'$ (see also Eq. (\ref{eq_model_1})). Here, since our goal is to have a framework that can effectively cope with the DSL problem in large-scale networks, we choose to propose the following approach. Let $\mathcal{O}''$ be the effective periphery node set of $\mathcal{V}_c''$ defined as
\begin{equation}\label{eq_percolation_rob_13}
    \mathcal{O}'' = \{u, \forall u\in \mathcal{O}, \exists \Gamma(u) \cap \mathcal{V}_c'' \neq \emptyset, t_u \neq \infty\},
\end{equation}
where
\begin{equation}\label{eq_percolation_rob_14}
    \mathcal{V}_c'' =  \bigcap_{\forall u \in \mathcal{O}'} \alpha(u).
\end{equation}
Letting $t_{\text{min}}=\argmin_{t_u} u, u\in \mathcal{O}''$, we first refine the time stamp by $t_u'=t_u-t_{\text{min}}$. Then, a Reverse-Influence-Sampling (RIS) \cite{borgs2014maximizing} like strategy is conducted to infer the source $\widehat{v}_c$, which works in the following processes. 1) Let $\Lambda=\{\}$ and $G''(\mathcal{V}', \mathcal{E}'')$ be the reverse network of $G'(\mathcal{V}', \mathcal{E}')$ satisfying that $|\mathcal{E}''| \equiv |\mathcal{E}'|$ and $e_{uv}\in \mathcal{E}''$ if $e_{vu}\in \mathcal{E}'$. 2) Randomly choose a node $u \in \mathcal{O}''$ and let $t_u''=t_0'+t_u'$, where $t_0'$ is randomly generated from $[0, \widehat{t}_0]$ and $\widehat{t}_0$ is a given parameter. 3) View $u$ as the source and transmits $\varsigma$ to one of its randomly chosen neighbors, and then it recovers. 4) Such transmission repeats $t_u''$ steps and denote the latest infected node by $v$. 5) Let $\Lambda =\Lambda \cup \{v\}$. 6) Repeat 2)-5) $T_\Lambda$ times. Using $\theta(v)$ to represent the frequency that a node $v \in \Lambda$ has regarding $\Lambda$, then we estimate the source $\widehat{v}_c$ by
\begin{equation}\label{eq_percolation_rob_15}
    \widehat{v}_c= \argmax_v \theta(v).
\end{equation}
The candidate set $\mathcal{V}_c$ (see also Eq. (\ref{eq_model_1})) is finally acquired by simply considering a few layers of neighbors of $\widehat{v}_c$.

\begin{remark}
Relying on AEF a finite $\mathcal{V}_c'$ can be achieved by a small $\mathcal{O}$, especially when $\mathcal{O}_d$ is large, i.e., the larger the $R_d$, the better the corresponding result, where $R_d=|\mathcal{O}_d|/|\mathcal{O}|$ characterizes the rate of $\mathcal{O}_d$ regarding $\mathcal{O}$. In particular, in tandem with the approach that we present to draw $\mathcal{V}_c$ from $\mathcal{V}_c'$, we name such framework as the \textbf{P}e\textbf{r}colation-based \textbf{E}volutionary \textbf{F}ramework (\textbf{PrEF}) for the diffusion-source-localization problem. Note that other strategies can also be further developed or integrated into PrEF to acquire $\mathcal{V}_c'$ based on $\mathcal{V}_c$, such as those existing DSL methods. {\hfill $\blacksquare$\par}
\end{remark}

\begin{figure}[htb]
  \centering
  \subfloat[]{
    \label{fig_re_1_1:a} 
    \includegraphics[width=0.24\linewidth]{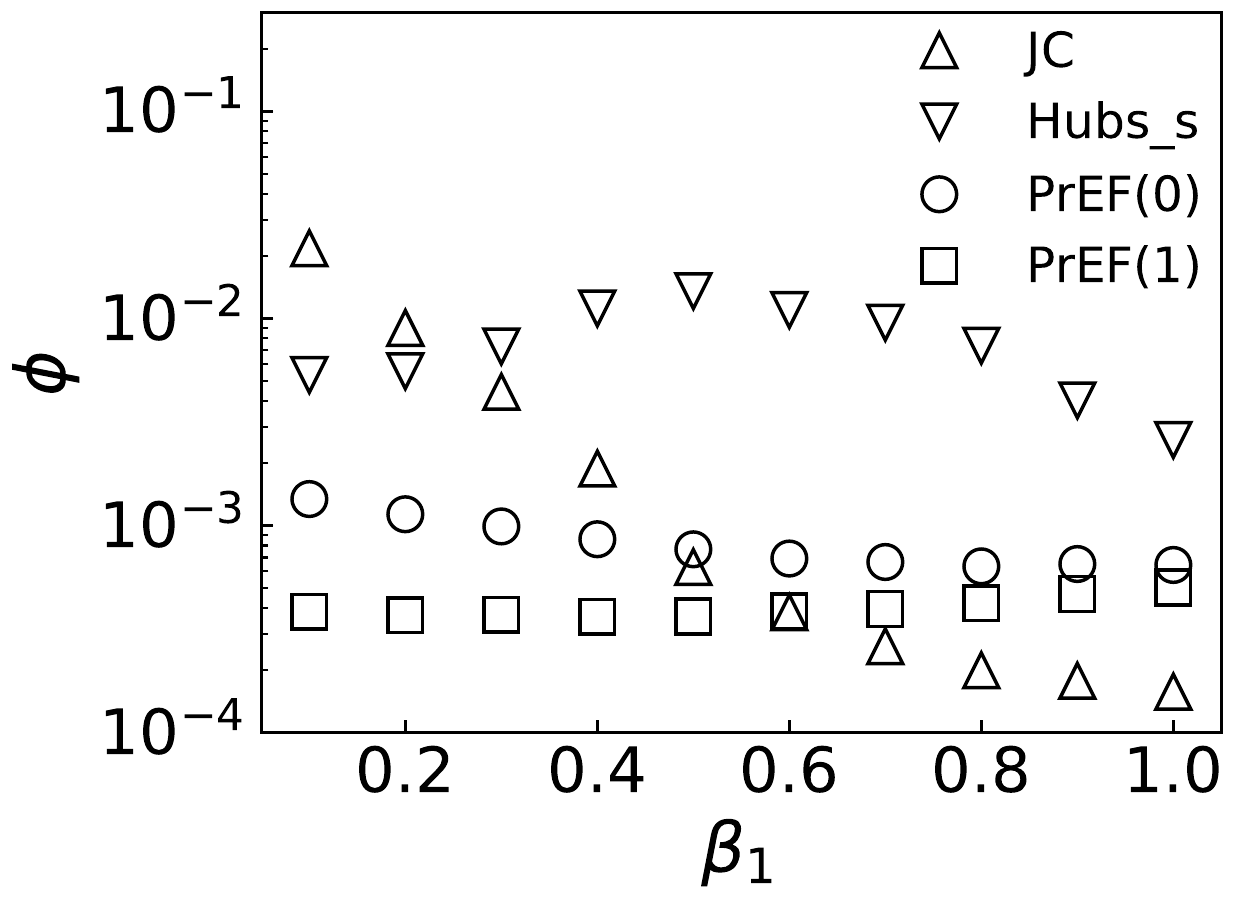}}
  \subfloat[]{
    \label{fig_re_1_1:b} 
    \includegraphics[width=0.24\linewidth]{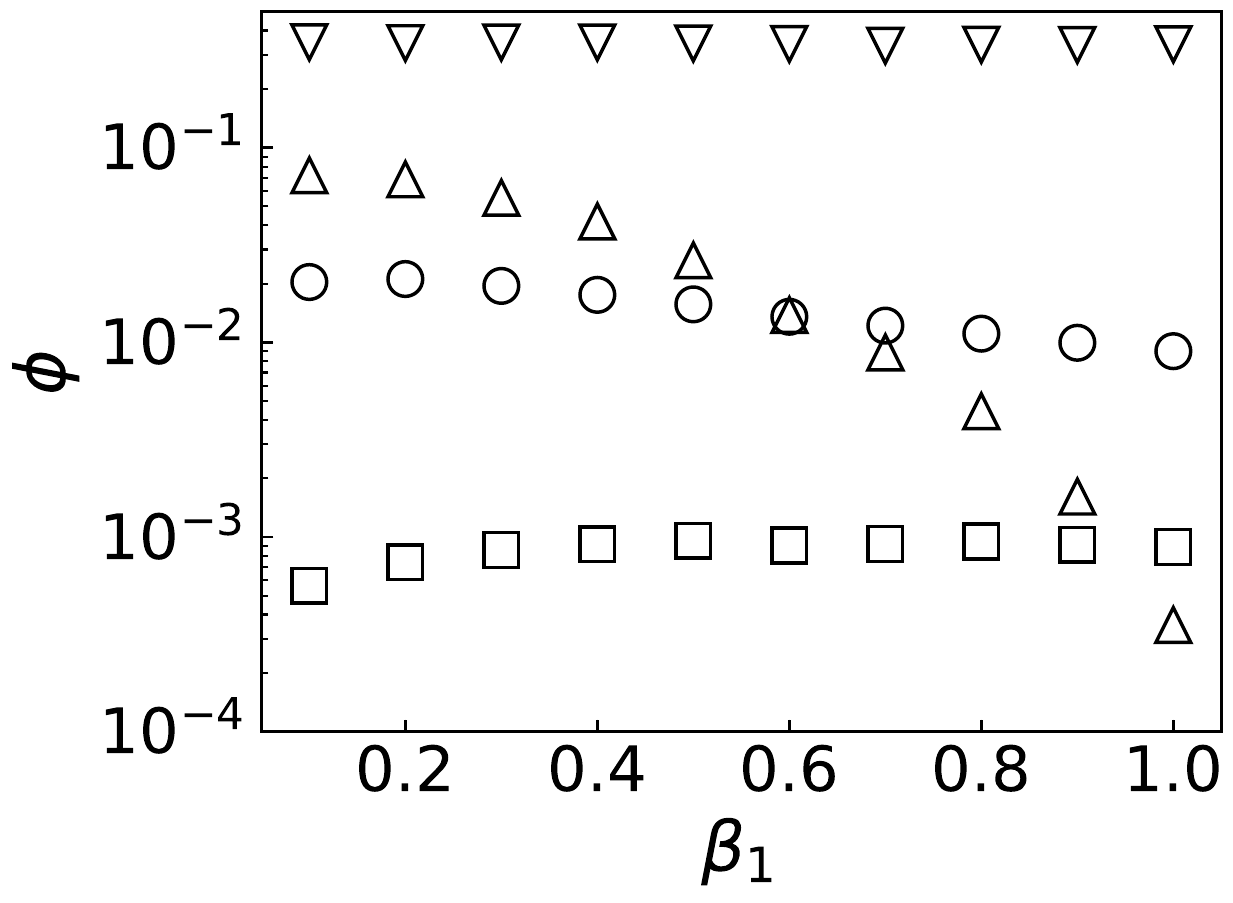}}
  \subfloat[]{
    \label{fig_re_1_1:c} 
    \includegraphics[width=0.24\linewidth]{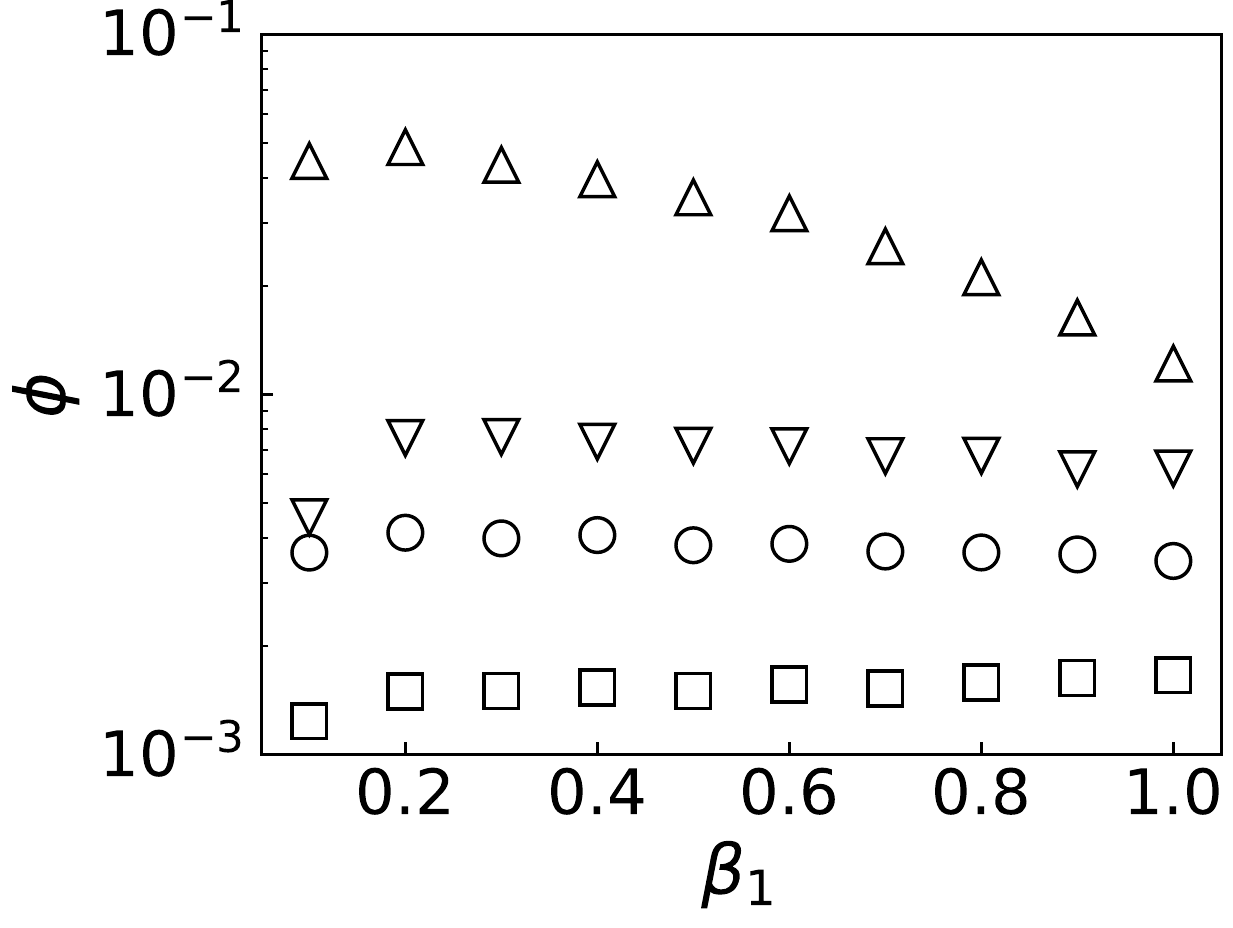}}
    \subfloat[]{
    \label{fig_re_1_1:d} 
    \includegraphics[width=0.24\linewidth]{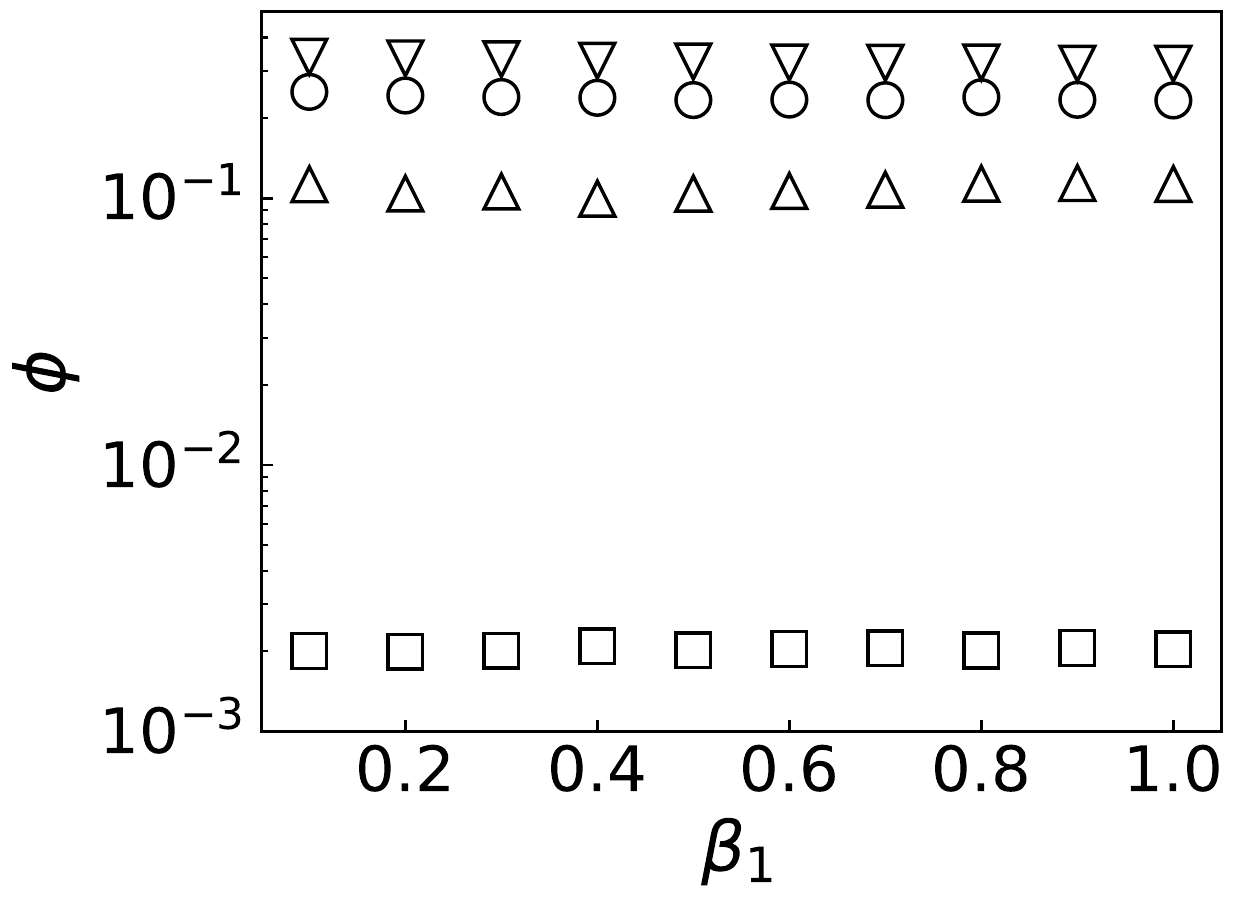}}
  \caption[]{The fraction of the candidate set $\phi$ as a function of the infection probability $\beta_1$ regarding SIR$_1$ with $\gamma_1=0.1$. (a) The ER network with $\langle k\rangle=3.50$ and $q_c=0.2100$. (b) The SF network with $\langle k\rangle=4.00$, $\ell=3.0$, and $q_c=0.1150$. (c) The PG network with $q_c=0.0810$. (d) The SCM network with $q_c=0.0664$. Samples are generated at $\varepsilon=0.10$.}
  \label{fig_re_1_1}
\end{figure}

\section{Results}

\textbf{Competitors.} We mainly compare the proposed approach with the Jordan-Center (JC) method \cite{zhu2014information} that generally achieves comparable results in most cases \cite{lokhov2014inferring, choi2020information, chai2021information}. For JC, the corresponding candidate set $\mathcal{V}_c$ is constructed based on the associated node rank since neighbor-based strategy usually results in much larger $\mathcal{V}_c$. Meanwhile, $\mathcal{O}$ consists of hubs is also considered as a baseline, say Hubs\_s. Besides, since most current DSL approaches do not work for large networks, we also verify the performance of the proposed method by comparing it with approaches from the network immunization field, including the collective influence (CI) \cite{morone2015influence}, the min-sum and reverse-greedy (MSRG) strategy \cite{braunstein2016network}, and the FINDER (FInding key players in Networks through DEep Reinforcement learning) method \cite{fan2020finding}.

\textbf{Settings.} JC considers all infected and recovered nodes to achieve the source localization. PrEF is conducted with $\widehat{\Delta}=50$, $\widehat{r}=1$, $\widehat{T}=20$, $\pi_1=1$, $\pi_2=\lfloor 0.1 \times n\rfloor$, $T_p=5,000$ for networks of $n\leqslant 10^5$, $T_p=2,500$ for $10^5<n\leqslant 10^6$, $T_p=500$ for $n > 10^6$ (AEF), and $T_\Lambda = 10^6$ (RIS). Besides, we use $\text{PrEF}(R_d)$ to represent PrEF regarding a specific $R_d$. In addition, for each network, $q_c$ is obtained at $\mathcal{G}(q_c)\thickapprox 0.005$ of AEF.


\textbf{Diffusion models.} SIR$_1$: $\beta_{uv}=\beta_1$ and $\gamma_{u}=\gamma_1$, $\forall u, v \in \mathcal{V}$. SIR$_2$: $\beta_{uv} \in [\beta_0, \beta_1]$ and $\gamma_{u}=0$, $\forall u, v \in \mathcal{V}$, i.e., the Susceptible-Infected (SI) model \cite{keeling2011modeling}. SIR$_3$: $\beta_{uv}\in [\beta_0, \beta_1]$ and $\gamma_{u}=1$, $\forall u, v \in \mathcal{V}$, i.e., the Independent Cascade (IC) model \cite{goldenberg2001talk, kempe2003maximizing}.

Letting $n(t, \text{I})$ and $n(t, \text{R})$ accordingly be the number of nodes in infection and recovery states at $t$ of a particular diffusion $\zeta(G, v_s, M, t)$, we generate a DSL sample by the following processes. 1) A node $v_s \in \mathcal{V}$ is randomly chosen as the diffusion source to trigger $\zeta$. 2) $\zeta$ is terminated at the moment when
$$(n(t, \text{I})+n(t, \text{R}))/n \geqslant \varepsilon,$$
where $\varepsilon$ is the outbreak range rate. Note that $(n(t, \text{I})+n(t, \text{R}))/n$ might be much larger than $\varepsilon$ if the infection probability is large.

\textbf{Evaluation metric.} We mainly consider the fraction of the candidate set (see also Eq. (\ref{eq_model_1})), $\phi$, to evaluate the performance of the proposed method, which is defined as
$$\phi= |\mathcal{V}_c|/n.$$
In what follows, $\phi$ is the mean drawn from over $1,000$ independent realizations if there is no special explantation. Besides, we also use $\phi(\cdot)$ to denote $\phi$ of a specific approach, such as $\phi(\text{PrEF})$ represents $\phi$ of PrEF.

\begin{table}[!htb]
\center
\caption[]{Experimental networks.}
\label{table_info_networks}
\begin{tabular}[c]{lrr}
\toprule
Networks & $n$ & $m$\\
\midrule
ER & $10,000$ & $35,000$	\\
SF & $10,000$ & $40,000$	\\
PG & $4,941$ & $6,594$	\\
SCM & $7,228$ & $24,784$	\\
LOCG & $196,591$ & $950,327$	\\
WG & $875,713$ & $4,322,051$\\
\bottomrule
\end{tabular}
\end{table}


\textbf{Networks.} We consider both model networks (including the ER network \cite{erdds1959random} and scale-free (SF) network \cite{barabasi1999emergence}) and empirical networks (including the Power Grid (PG) network \cite{watts1998collective}, the Scottish-cattle-movement (SCM) network \cite{clusella2016immunization}, the loc-Gowalla (LOCG) network \cite{cho2011friendship}, and the web-Google (WG) network \cite{leskovec2009community}). We choose the PG network since it is widely used to evaluate the performance of DSL approaches. Rather than that, the rest are all highly associated with the DSL problem. In particular, the SCM network is a network of Scottish cattle movements, on which the study of the DSL problem plays important role for food security \cite{keeling2003modelling}. Besides, the LOCG is a location-based online social network and the WG network is a network of Google web whose nodes represent web pages and edges are hyperlinks among them. The study of them can potentially contribute to the containment of misinformation. The basic information regarding those networks can be found in Table \ref{table_info_networks}.

\begin{figure}[htb]
  \centering
  \subfloat[]{
    \label{fig_re_1_2:a} 
    \includegraphics[width=0.24\linewidth]{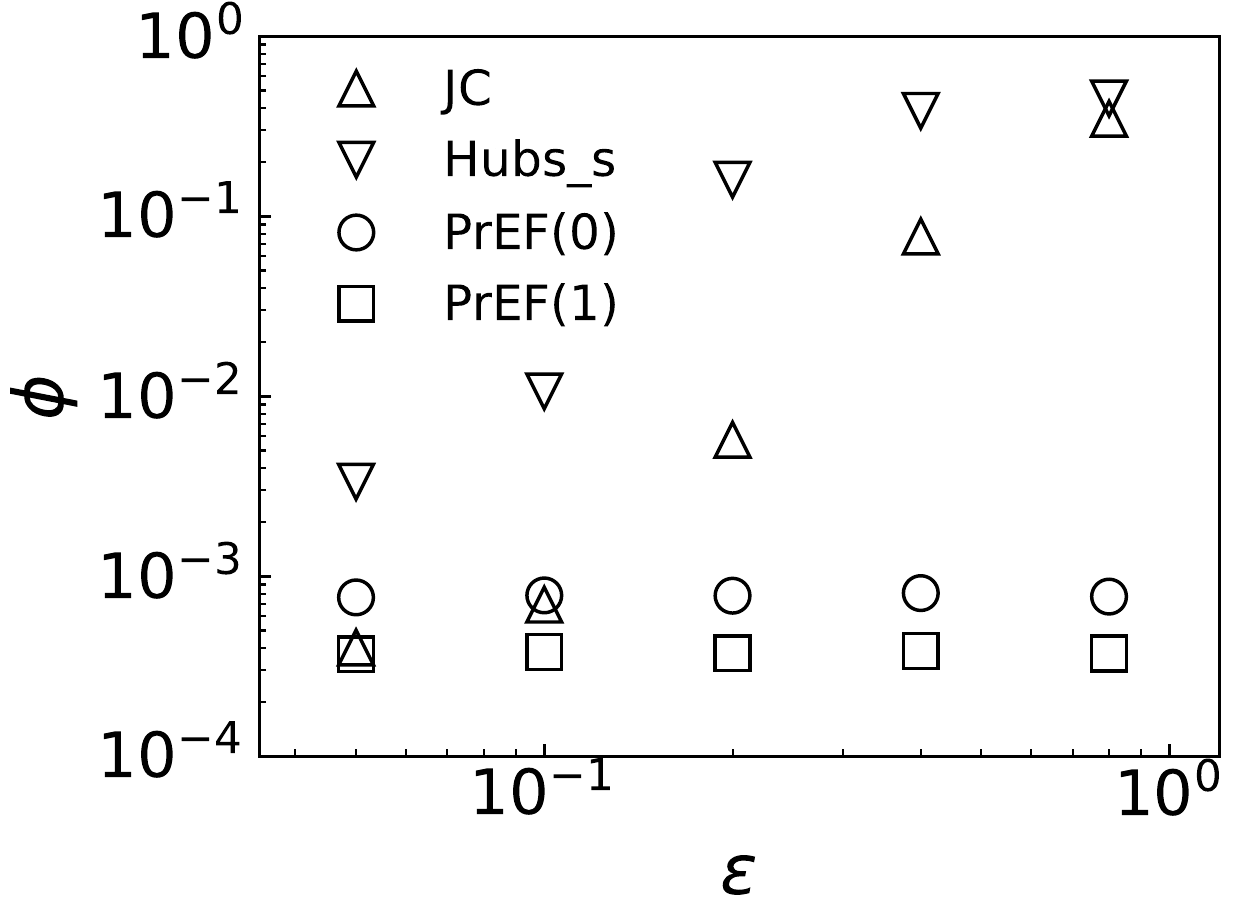}}
  \subfloat[]{
    \label{fig_re_1_2:b} 
    \includegraphics[width=0.24\linewidth]{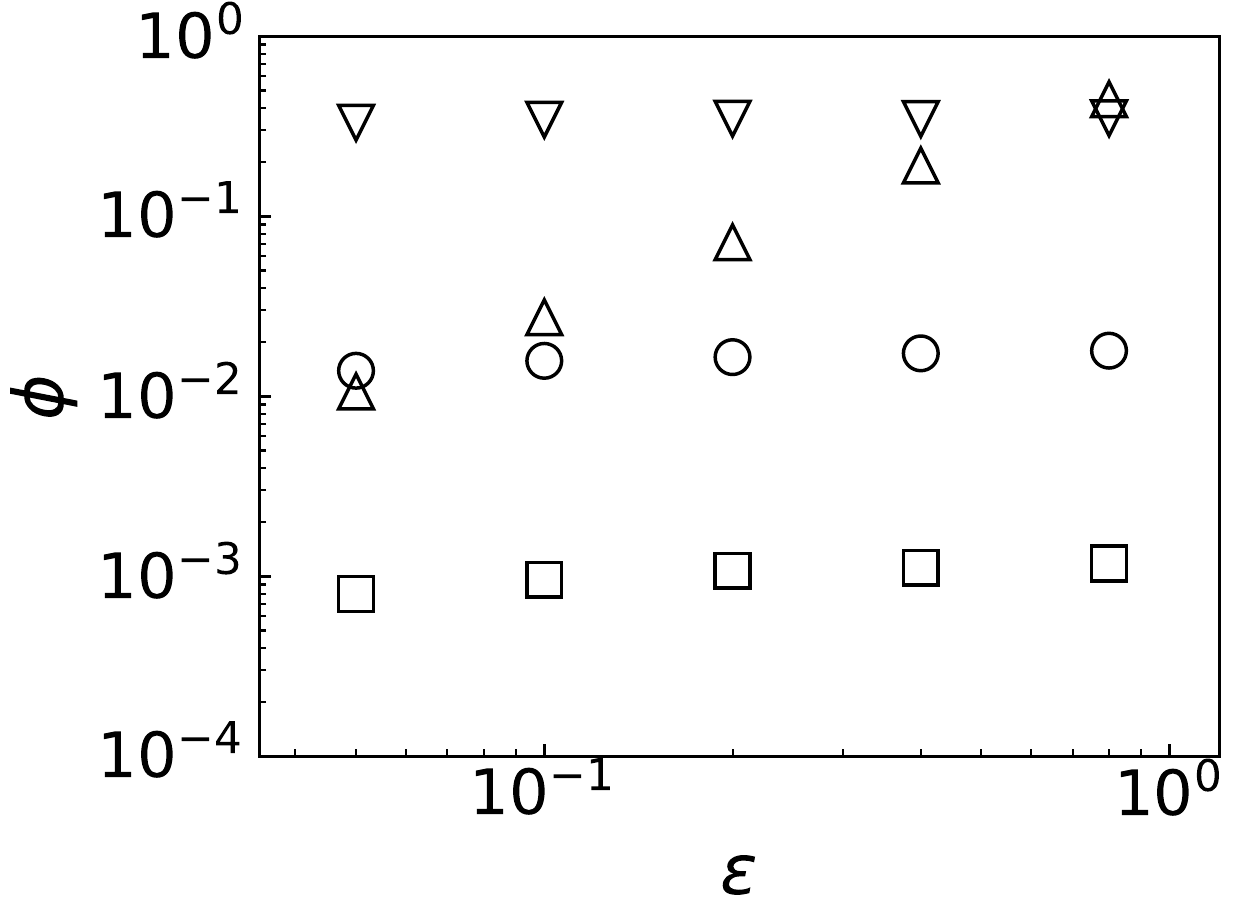}}
  \subfloat[]{
    \label{fig_re_1_2:c} 
    \includegraphics[width=0.24\linewidth]{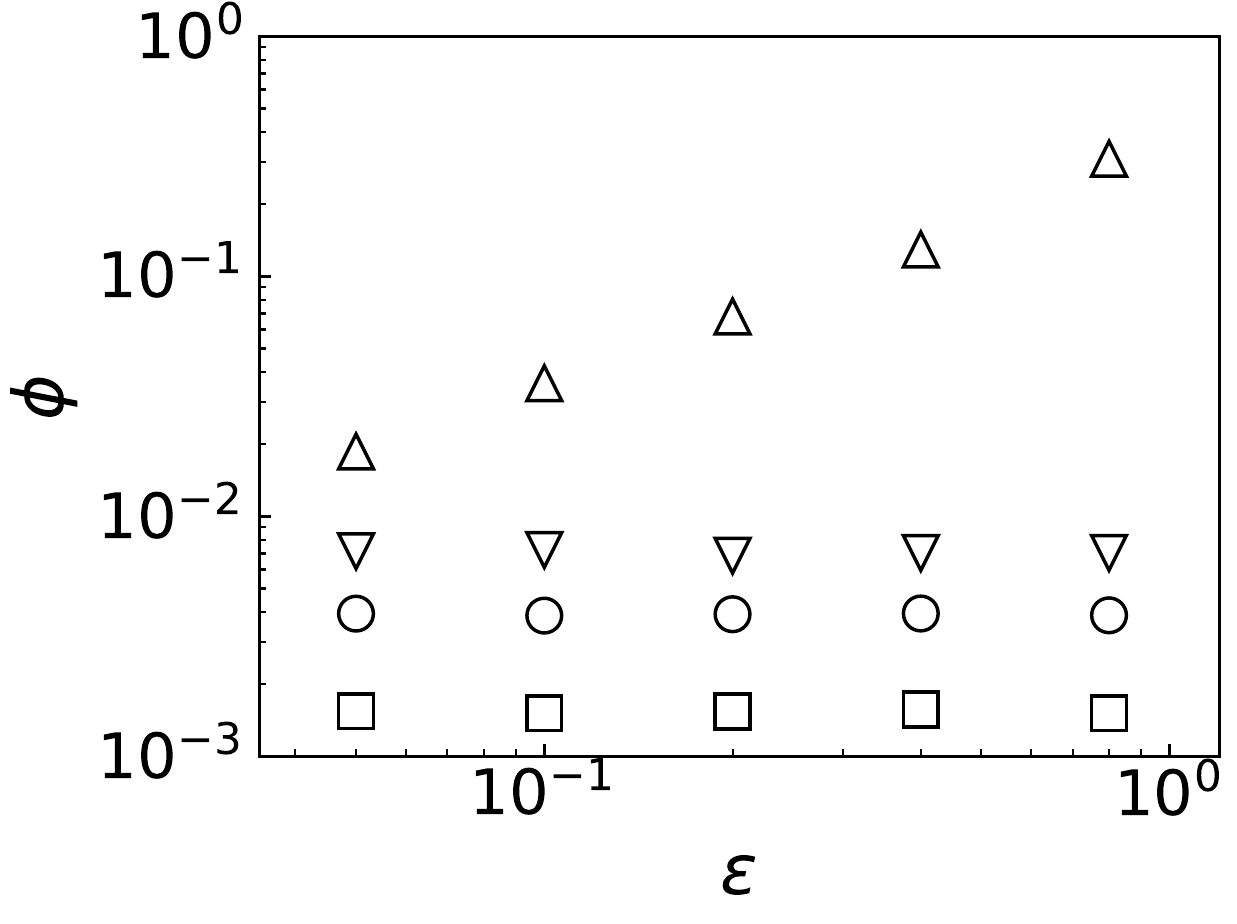}}
    \subfloat[]{
    \label{fig_re_1_2:d} 
    \includegraphics[width=0.24\linewidth]{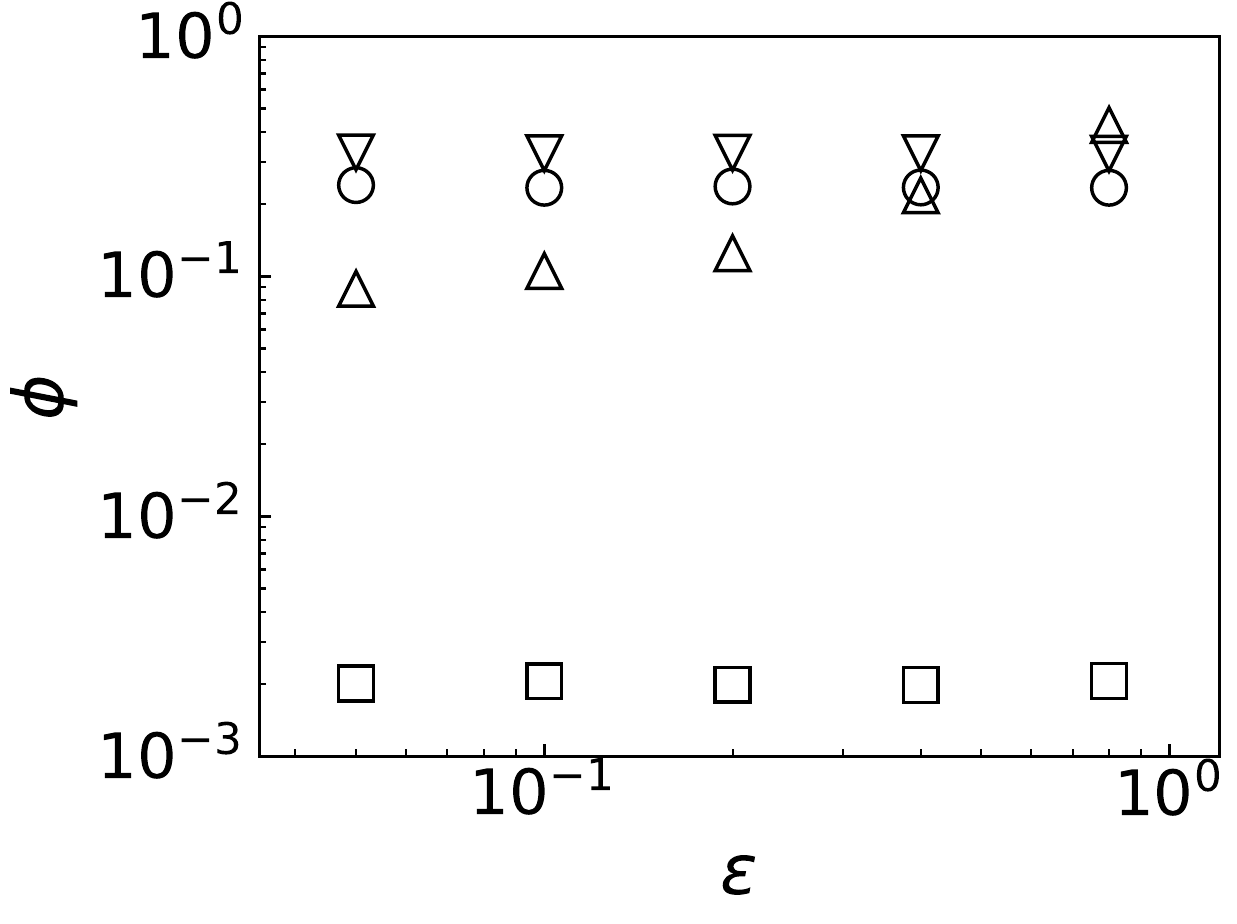}}
  \caption[]{$\phi$ as a function of the outbreak range rate $\varepsilon$, where SIR$_1$ with $\beta_1=0.5$ and $\gamma_1=0.1$ is considered. (a) The ER network. (b) The SF network. (c) The PG network. (d) The SCM network.}
  \label{fig_re_1_2}
\end{figure}


\begin{figure}[htb]
  \centering
  \subfloat[]{
    \label{fig_re_3_3:a} 
    \includegraphics[width=0.24\linewidth]{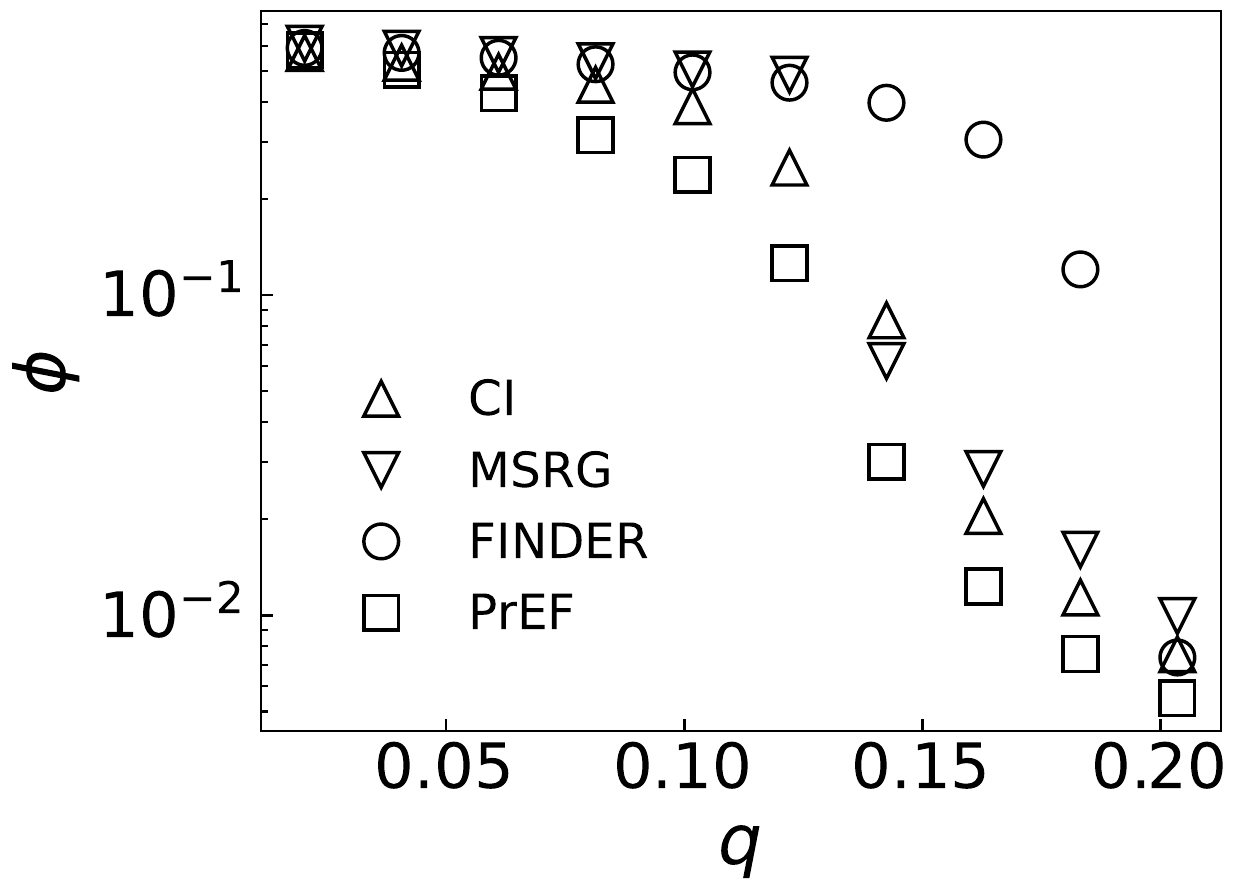}}
  \subfloat[]{
    \label{fig_re_3_3:b} 
    \includegraphics[width=0.24\linewidth]{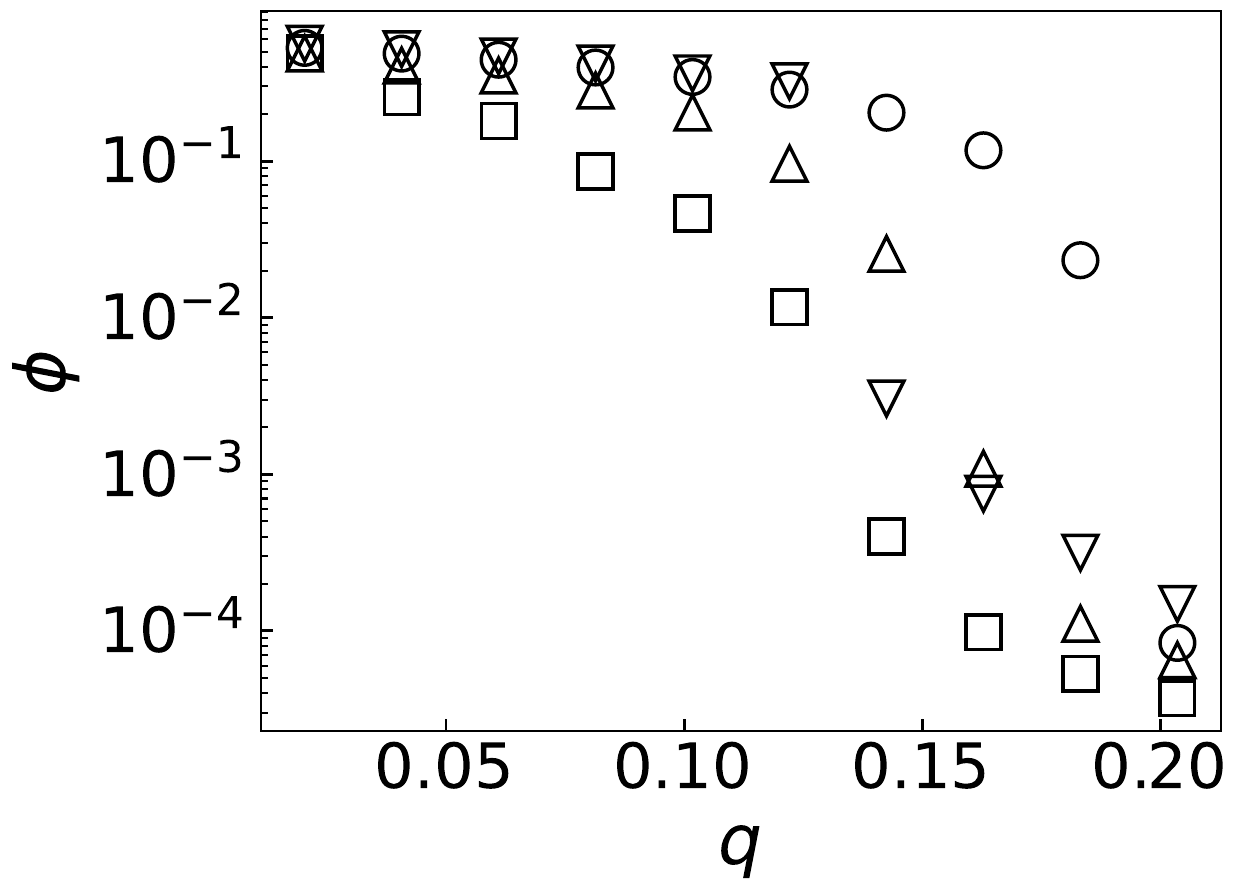}}
  \subfloat[]{
    \label{fig_re_3_3:c} 
    \includegraphics[width=0.24\linewidth]{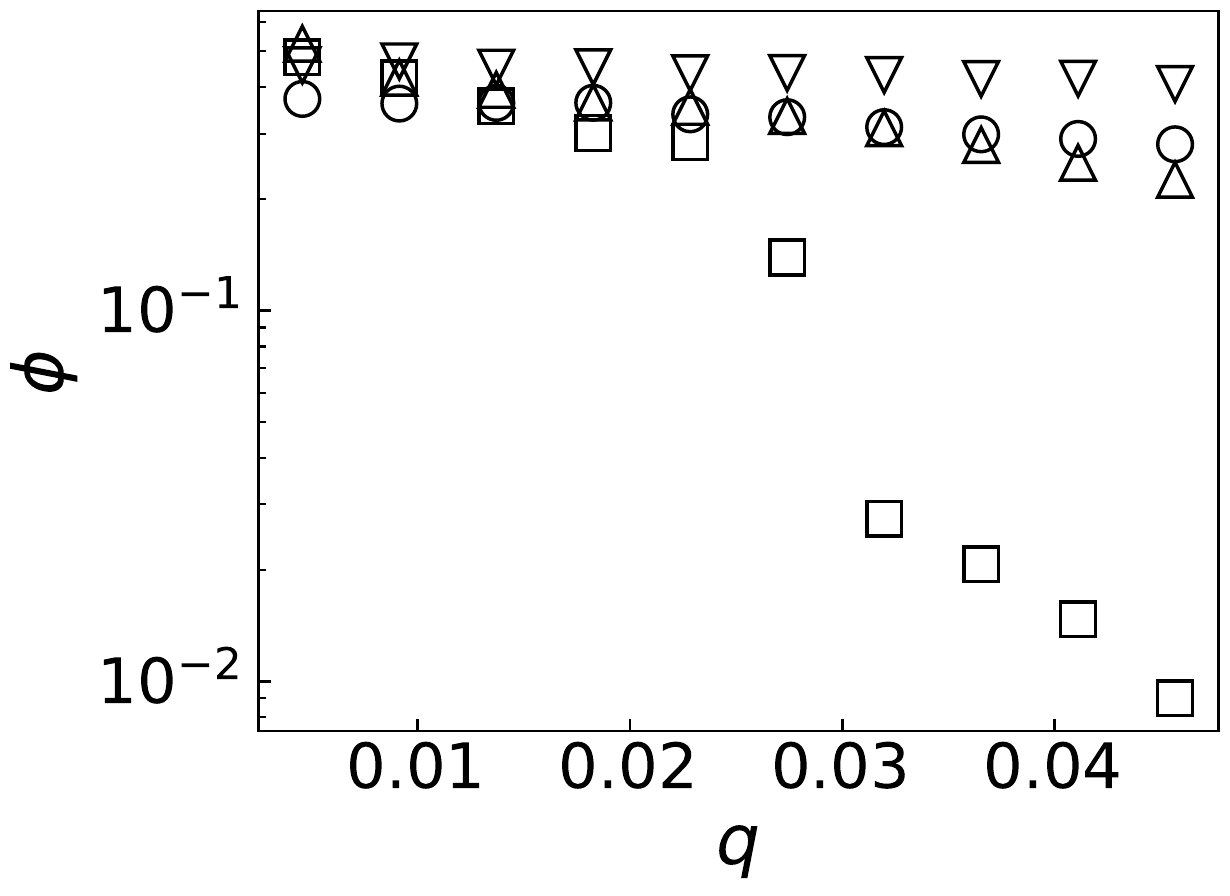}}
    \subfloat[]{
    \label{fig_re_3_3:d} 
    \includegraphics[width=0.24\linewidth]{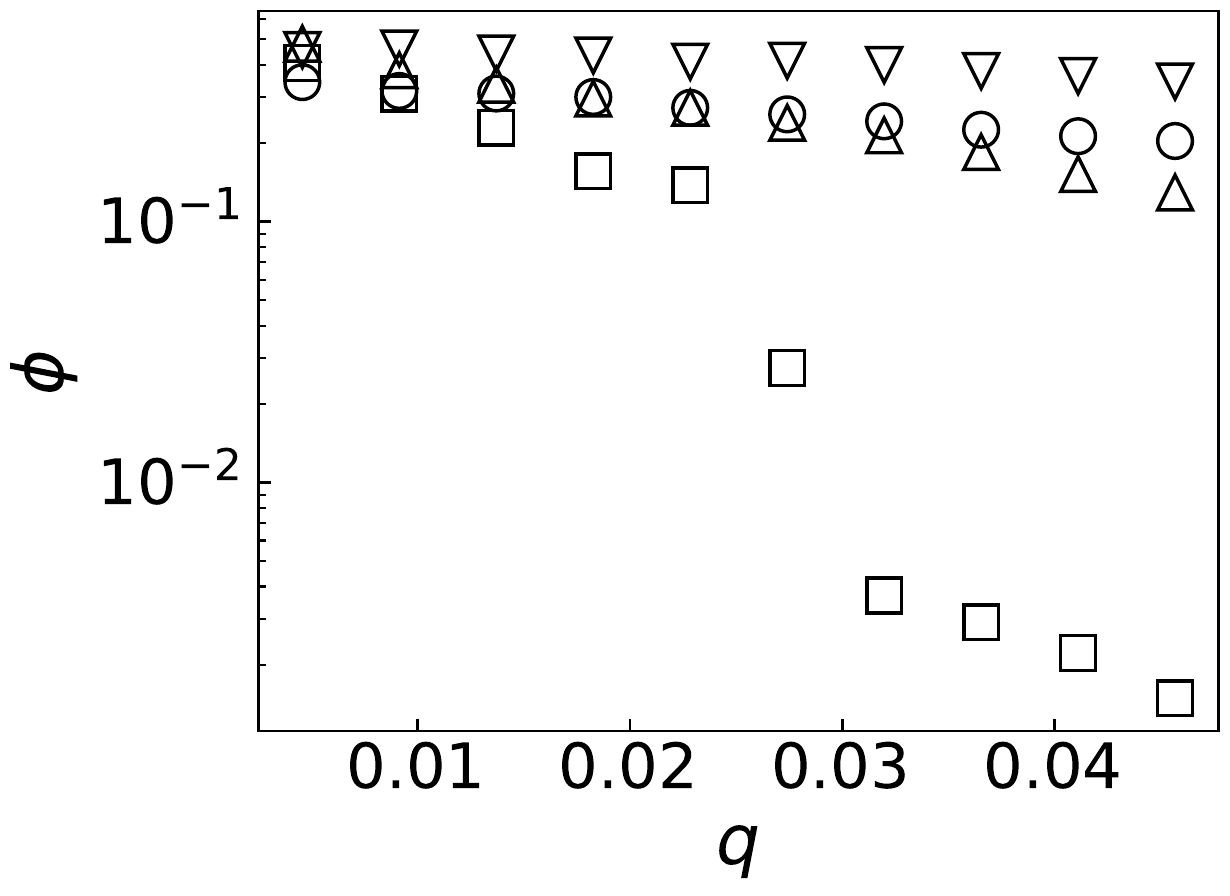}}
  \caption[]{$\phi$ as a function of the fraction of observers $q$. Here SIR$_1$ is conducted with $\beta_1=0.5$, $\gamma_1=0.1$, and $\varepsilon=0.10$. (a) and (b) The LOCG network. (c) and (d) The WG network. Besides, $R_d=0$ is for (a) and (c), and $R_d=1$ for (b) and (d).}
  \label{fig_re_3_3}
\end{figure}


\textbf{Results.} We first fix $q$ to verify the performance of PrEF over varied infection probability $\beta_1$. Indeed, if the diffusion is symmetrical, JC would be an effective estimator (see Figs. \ref{fig_re_1_1:a} and \ref{fig_re_1_1:b} when $\beta_1$ is large). But such effectiveness sharply decreases as $\beta_1$ decreases. By contrast, PrEF has steady performance for the whole range of $\beta_1$ and is much better than JC when $\beta_1$ is small, such as $\phi(\text{PrEF}(1))=0.0004$ versus $\phi(\text{JC})=0.0721$ at $\beta_1=0.1$ in the SF network. Besides, $\text{PrEF}(0)$ apparently works better in the ER network compared to the SF network, which indicates that $k_\text{max}$ might have impact on the effectiveness of $\text{PrEF}(0)$ since the SF network has a much larger $k_\text{max}$. To further demonstrate that, we also consider two empirical networks: the Power Grid network (with $k_\text{max}=19$) and the Scottish network (with $k_\text{max}=3667$). As shown in Figs. \ref{fig_re_1_2:c} and \ref{fig_re_1_2:d}, even Hubs\_s is more effective than JC in the Power Grid network but, JC, Hubs\_s, and $\text{PrEF}(0)$ all fail in the Scottish network. Rather than that, $\text{PrEF}(1)$ works extremely well in both cases. Further considering the fraction of the candidate set $\phi$ as a function of the outbreak range rate $\varepsilon$ (Fig. \ref{fig_re_1_2}), $\text{PrEF}(0)$ and $\text{PrEF}(1)$ still have stable performance while $\phi(\text{JC})$ rapidly increases as $\phi$.


\begin{figure}[htb]
  \centering
  \subfloat[]{
    \label{fig_re_6_2:a} 
    \includegraphics[width=0.24\linewidth]{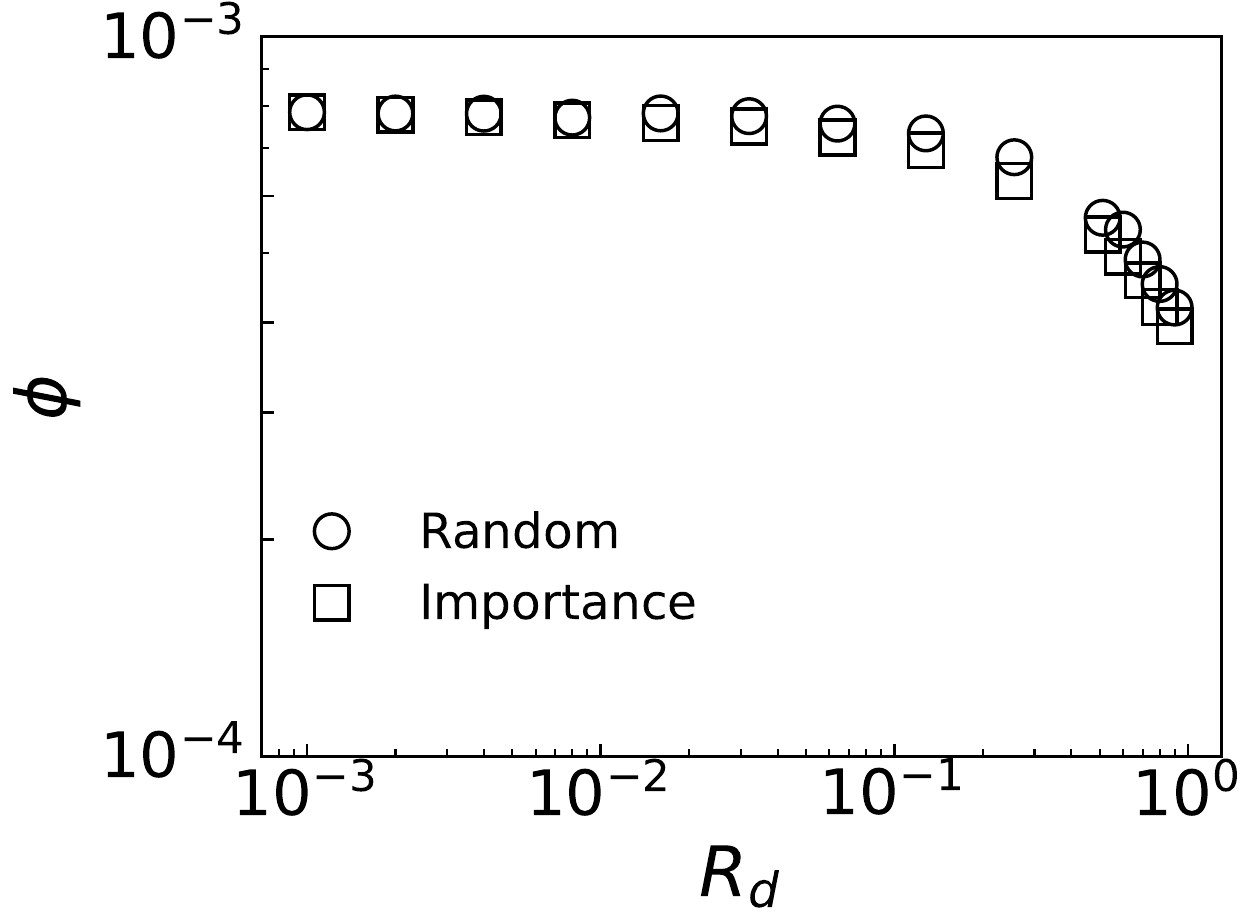}}
  \subfloat[]{
    \label{fig_re_6_2:b} 
    \includegraphics[width=0.24\linewidth]{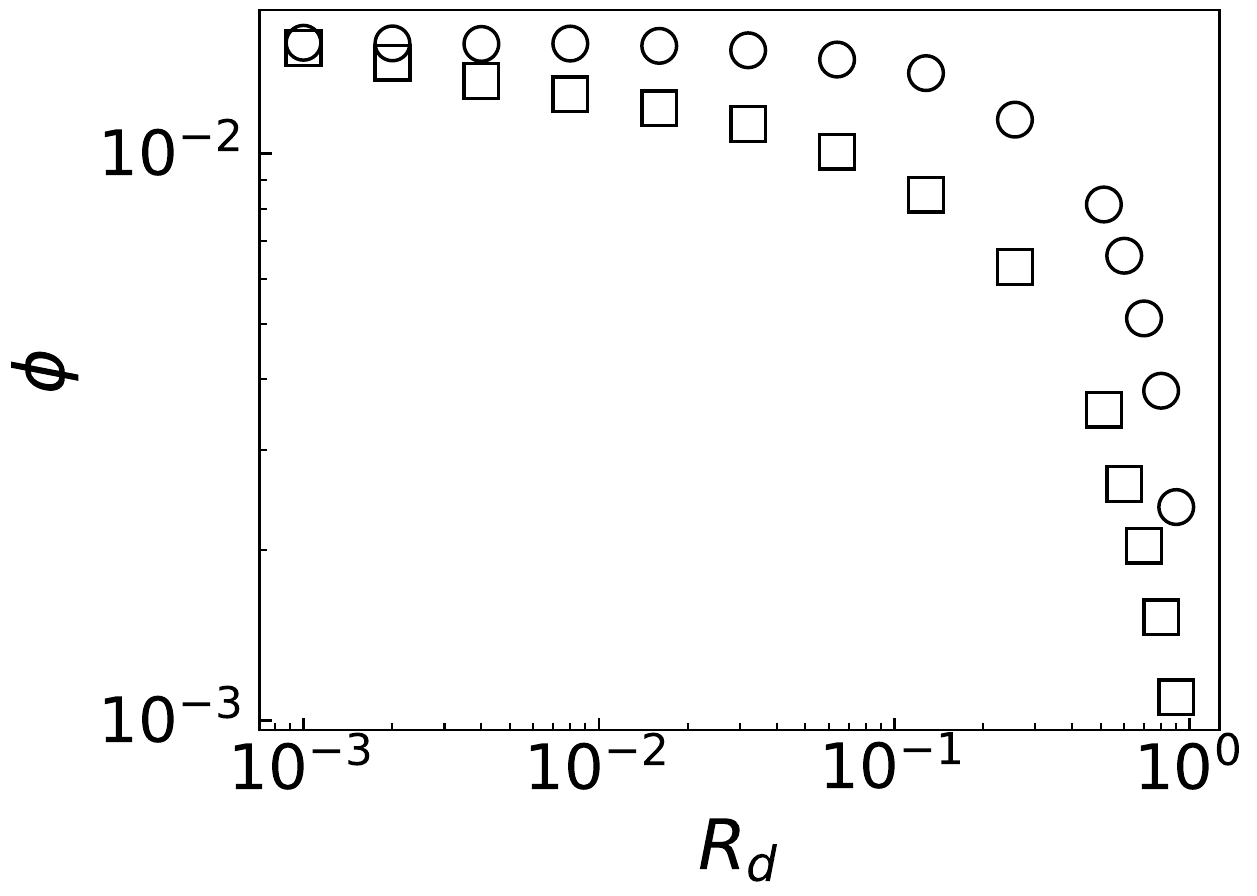}}
  \subfloat[]{
    \label{fig_re_6_2:c} 
    \includegraphics[width=0.24\linewidth]{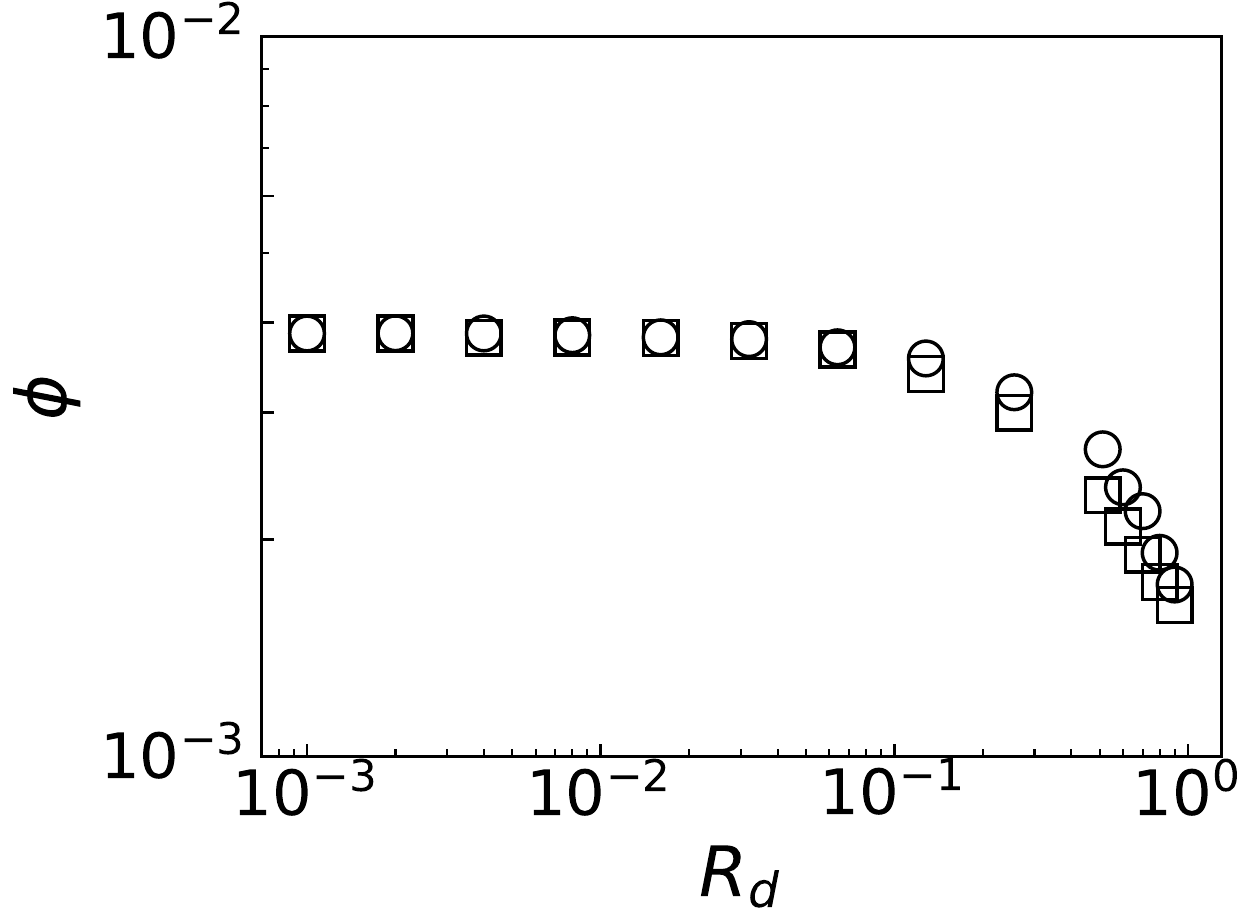}}
    \subfloat[]{
    \label{fig_re_6_2:d} 
    \includegraphics[width=0.24\linewidth]{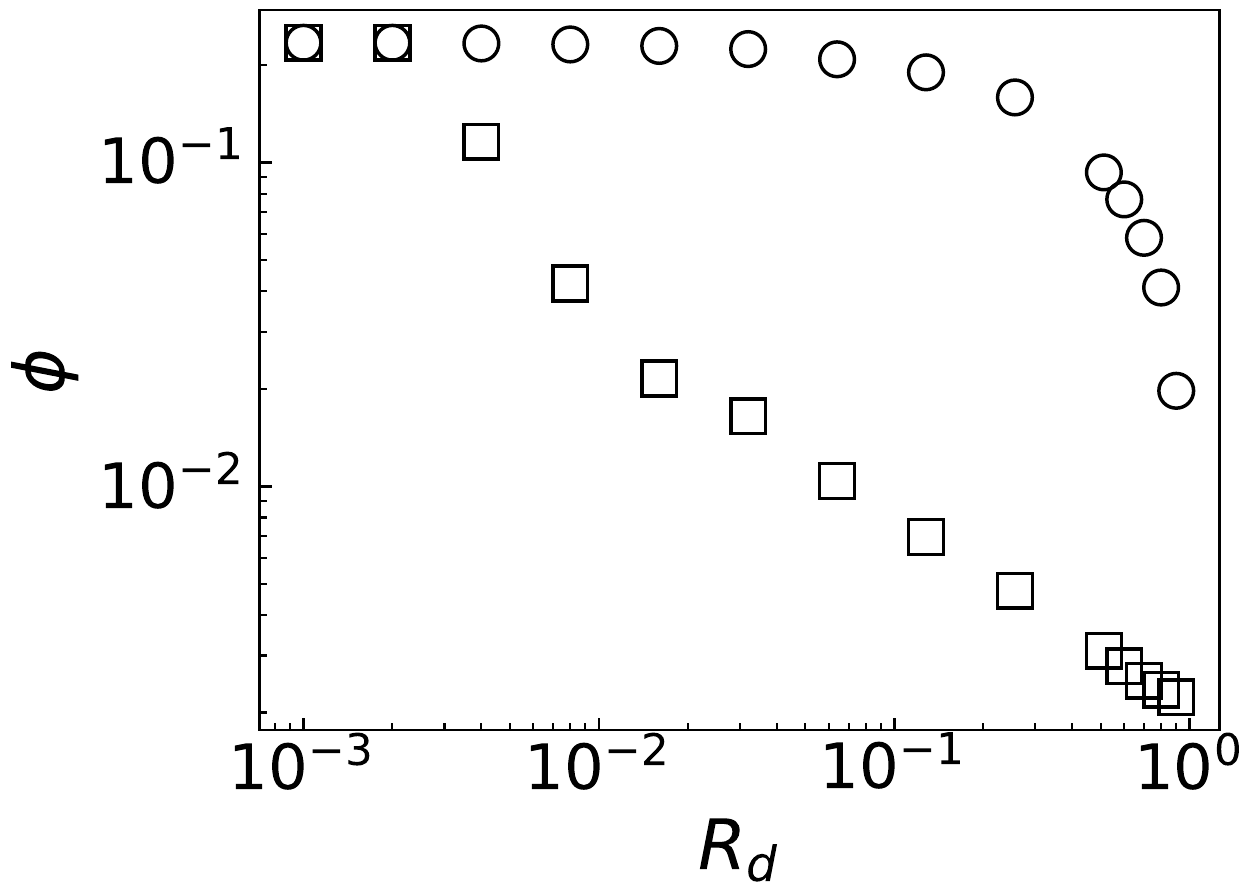}}
  \caption[]{$\phi$ of $R_d$ regarding SIR$_1$ with $\beta_1=0.5$ and $\gamma_1=0.1$, where `Random' represents that $\mathcal{O}_d$ is randomly chosen from $\mathcal{O}$ while `Importance' corresponds to the case that $\mathcal{O}_d$ is generated relying on $\mathcal{S}$. (a) The ER network. (b) The SF network. (c) The PG network. (d) The SCM network. Samples are generated at $\varepsilon=0.10$.}
  \label{fig_re_6_2}
\end{figure}


\begin{figure}[htb]
  \centering
  \subfloat[]{
    \label{fig_re_7_2:a} 
    \includegraphics[width=0.24\linewidth]{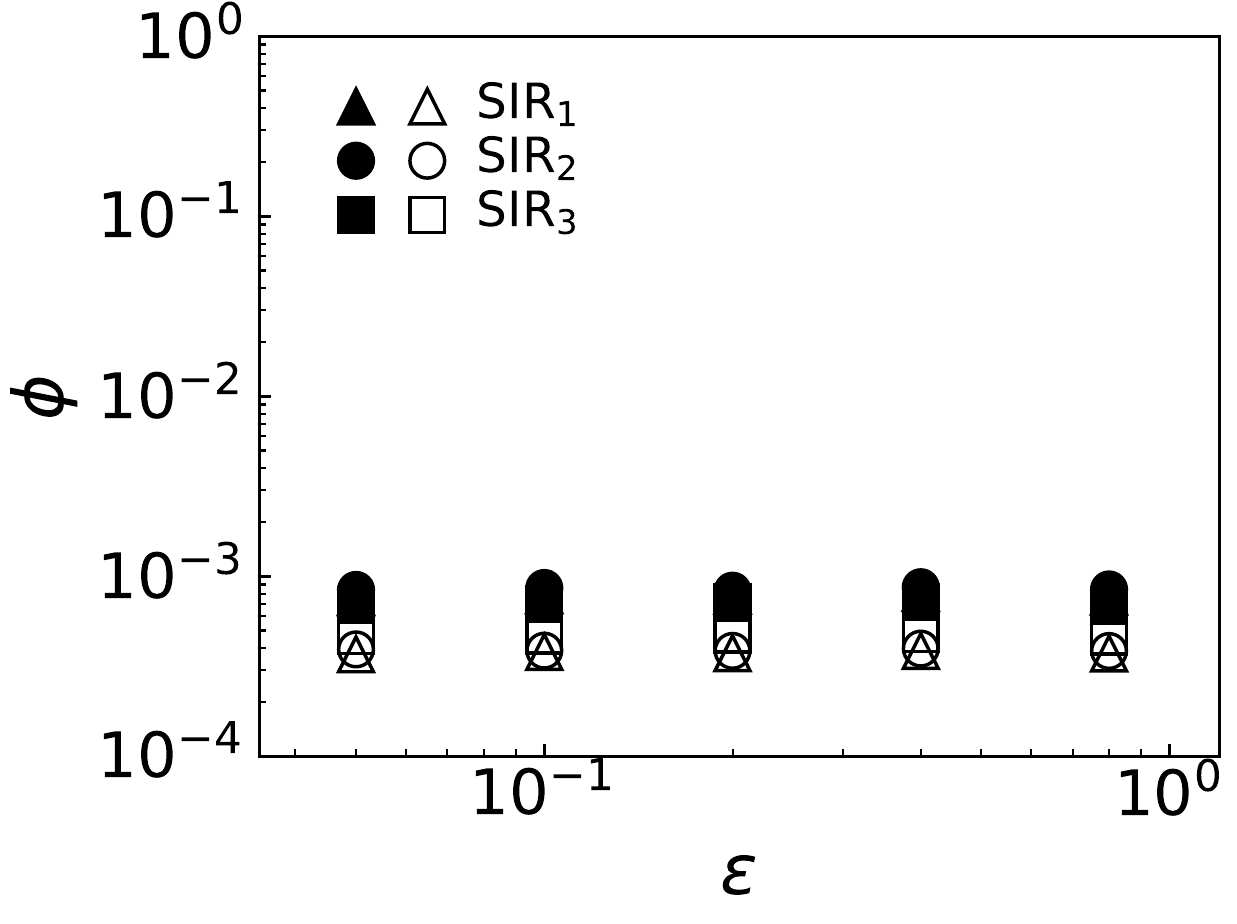}}
  \subfloat[]{
    \label{fig_re_7_2:b} 
    \includegraphics[width=0.24\linewidth]{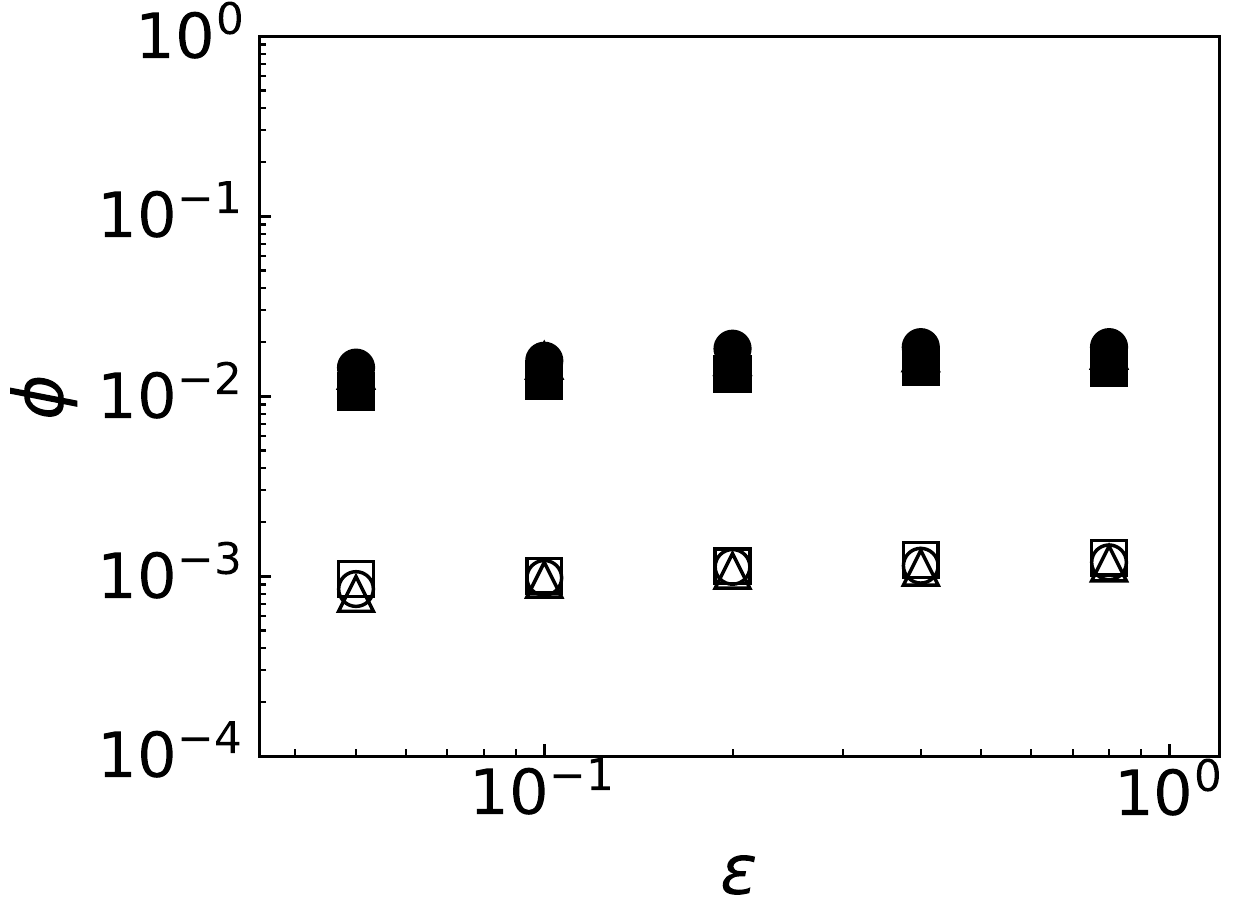}}
  \subfloat[]{
    \label{fig_re_7_2:c} 
    \includegraphics[width=0.24\linewidth]{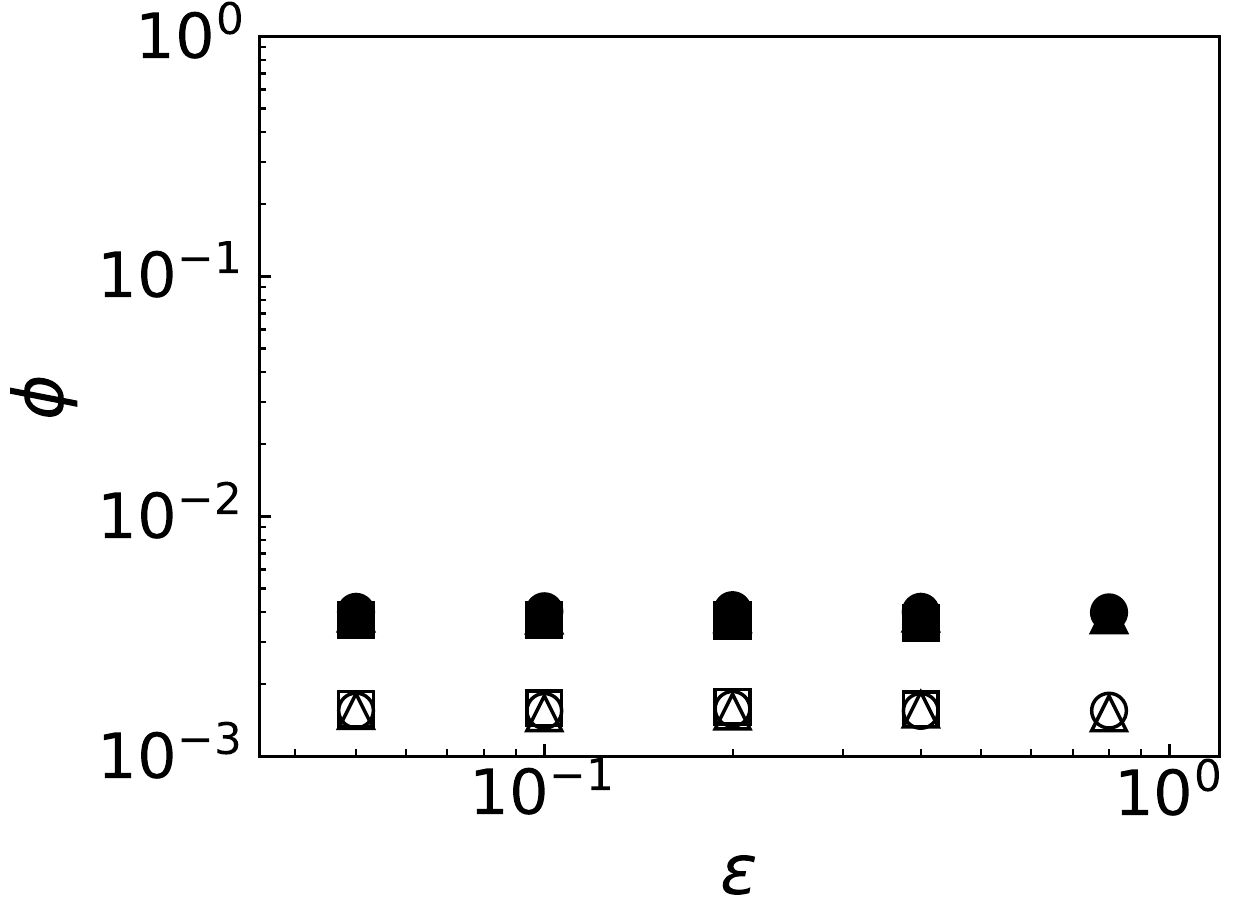}}
    \subfloat[]{
    \label{fig_re_7_2:d} 
    \includegraphics[width=0.24\linewidth]{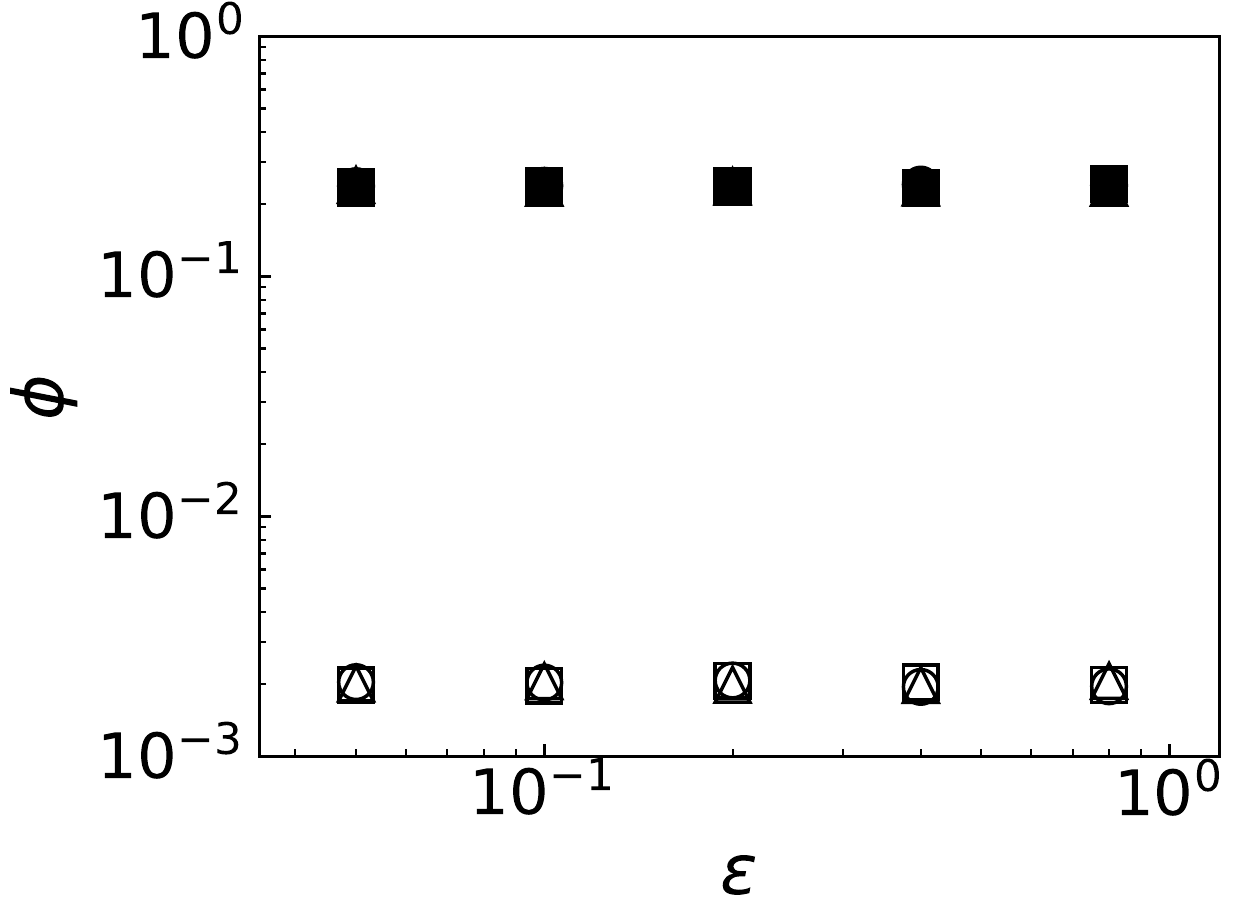}}
  \caption[]{$\phi$ of $\varepsilon$ regarding SIR$_1$, SIR$_2$, and SIR$_3$, where SIR$_1$ is with $\beta_1=0.5$ and $\gamma_1=0.1$, and $\beta_{uv} \in [0, 1]$ of SIR$_2$ and $\beta_{uv} \in [0.5, 1]$ of SIR$_3$ both are randomly generated. (a) The ER network. (b) The SF network. (c) The PG network. (d) The SCM network. Solid and unfilled marks are associated with $\text{PrEF}(0)$ and $\text{PrEF}(1)$, respectively.}
  \label{fig_re_7_2}
\end{figure}


We further evaluate the performance of $\text{PrEF}$ under different $q$ by comparing it with CI, MSRG, and FINDER on the two large networks. From those results shown in Fig. \ref{fig_re_3_3}, we have the following conclusions: i) $\phi \rightarrow 1$ when $q \rightarrow 0$, which is in accordance with our previous discussion, i.e., $\mathcal{G}(q) \rightarrow 1$ when $q \rightarrow 0$; ii) A specific method $\mathcal{S}$ that has better performance regarding $R_d=0$ usually also works better with respect to $R_d=1$; iii) For a specific $q$, PrEF always has much smaller $\phi$ compared to CI, MSRG, and FINDER, especially for the WG network. Indeed, the size of the observer set $\mathcal{O}$, the value of $R_d$ (see also Fig. \ref{fig_re_6_2}), the strategy generating $\mathcal{O}$ all play fundamental roles for minimizing $\phi$. In particular, $\text{PrEF}$ has the best performance for almost all range of $q$. Besides, results in Fig. \ref{fig_re_7_2} further demonstrate that the proposed method is also stable against varied diffusion models.

\section{Conclusion}

Aiming at the development of the-state-of-the-art approach to cope with the DSL problem for large networks, the PrEF method has been developed based on the network percolation and evolutionary computation, which can effectively narrow our search region of the diffusion source. Specifically, We have found that the DSL problem is in a degree equivalent to the network immunization problem if viewing immune nodes as observers, and hence it can be tackled in a similar scheme. In particular, we have demonstrated that the search region would be bounded by the LCC if the direction information of the diffusion is known, regardless of the network structure. But for the case that only the time stamp is recorded, both LCC and the largest degree have impact on the search region. We have also conducted extensive experiments to evaluate the performance of the proposed method. Results show that our method is much more effective, efficient, and stable compared to existing approaches.

\bibliographystyle{IEEEbibsty}
\bibliography{refs}

\begin{thebibliography}{10}
\providecommand{\url}[1]{#1}
\csname url@samestyle\endcsname
\providecommand{\newblock}{\relax}
\providecommand{\bibinfo}[2]{#2}
\providecommand{\BIBentrySTDinterwordspacing}{\spaceskip=0pt\relax}
\providecommand{\BIBentryALTinterwordstretchfactor}{4}
\providecommand{\BIBentryALTinterwordspacing}{\spaceskip=\fontdimen2\font plus
\BIBentryALTinterwordstretchfactor\fontdimen3\font minus
  \fontdimen4\font\relax}
\providecommand{\BIBforeignlanguage}[2]{{%
\expandafter\ifx\csname l@#1\endcsname\relax
\typeout{** WARNING: IEEEtran.bst: No hyphenation pattern has been}%
\typeout{** loaded for the language `#1'. Using the pattern for}%
\typeout{** the default language instead.}%
\else
\language=\csname l@#1\endcsname
\fi
#2}}
\providecommand{\BIBdecl}{\relax}
\BIBdecl

\bibitem{shah2010detecting}
D.~Shah and T.~Zaman, ``Detecting sources of computer viruses in networks:
  theory and experiment,'' in \emph{Proceedings of the ACM SIGMETRICS
  international conference on Measurement and modeling of computer systems},
  2010, pp. 203--214.

\bibitem{pinto2012locating}
P.~C. Pinto, P.~Thiran, and M.~Vetterli, ``Locating the source of diffusion in
  large-scale networks,'' \emph{Physical review letters}, vol. 109, no.~6, p.
  068702, 2012.

\bibitem{ali2019epa}
S.~S. Ali, T.~Anwar, A.~Rastogi, and S.~A.~M. Rizvi, ``Epa: Exoneration and
  prominence based age for infection source identification,'' in
  \emph{Proceedings of the 28th ACM International Conference on Information and
  Knowledge Management}, 2019, pp. 891--900.

\bibitem{harvell2002climate}
C.~D. Harvell, C.~E. Mitchell, J.~R. Ward, S.~Altizer, A.~P. Dobson, R.~S.
  Ostfeld, and M.~D. Samuel, ``Climate warming and disease risks for
  terrestrial and marine biota,'' \emph{Science}, vol. 296, no. 5576, pp.
  2158--2162, 2002.

\bibitem{brauer2012mathematical}
F.~Brauer, C.~Castillo-Chavez, and C.~Castillo-Chavez, \emph{Mathematical
  models in population biology and epidemiology}.\hskip 1em plus 0.5em minus
  0.4em\relax Springer, 2012, vol.~2.

\bibitem{mcmichael2003climate}
A.~J. McMichael, D.~H. Campbell-Lendrum, C.~F. Corval{\'a}n, K.~L. Ebi,
  A.~Githeko, J.~D. Scheraga, and A.~Woodward, \emph{Climate change and human
  health: risks and responses}.\hskip 1em plus 0.5em minus 0.4em\relax World
  Health Organization, 2003.

\bibitem{jamison2013global}
D.~T. Jamison, L.~H. Summers, G.~Alleyne, K.~J. Arrow, S.~Berkley,
  A.~Binagwaho, F.~Bustreo, D.~Evans, R.~G. Feachem, J.~Frenk \emph{et~al.},
  ``Global health 2035: a world converging within a generation,'' \emph{The
  Lancet}, vol. 382, no. 9908, pp. 1898--1955, 2013.

\bibitem{zhou2020survey}
X.~Zhou and R.~Zafarani, ``A survey of fake news: Fundamental theories,
  detection methods, and opportunities,'' \emph{ACM Computing Surveys (CSUR)},
  vol.~53, no.~5, pp. 1--40, 2020.

\bibitem{sahoo2020demystifying}
S.~Sahoo, S.~K. Padhy, J.~Ipsita, A.~Mehra, and S.~Grover, ``Demystifying the
  myths about covid-19 infection and its societal importance,'' \emph{Asian
  journal of psychiatry}, vol.~54, p. 102244, 2020.

\bibitem{fakenewseconomy}
K.~Rapoza, ``Can 'fake news' impact the stock market?''
  \url{https://www.forbes.com/sites/kenrapoza/2017/02/26/can-fake-news-impact-the-stock-market/?sh=5f820b392fac},
  2017, accessed: 2021-09-15.

\bibitem{choi2020information}
J.~Choi, S.~Moon, J.~Woo, K.~Son, J.~Shin, and Y.~Yi, ``Information source
  finding in networks: Querying with budgets,'' \emph{IEEE/ACM Transactions on
  Networking}, vol.~28, no.~5, pp. 2271--2284, 2020.

\bibitem{zhu2014information}
K.~Zhu and L.~Ying, ``Information source detection in the sir model: A
  sample-path-based approach,'' \emph{IEEE/ACM Transactions on Networking},
  vol.~24, no.~1, pp. 408--421, 2014.

\bibitem{wang2014rumor}
Z.~Wang, W.~Dong, W.~Zhang, and C.~W. Tan, ``Rumor source detection with
  multiple observations: Fundamental limits and algorithms,'' \emph{ACM
  SIGMETRICS Performance Evaluation Review}, vol.~42, no.~1, pp. 1--13, 2014.

\bibitem{jiang2016identifying}
J.~Jiang, S.~Wen, S.~Yu, Y.~Xiang, and W.~Zhou, ``Identifying propagation
  sources in networks: State-of-the-art and comparative studies,'' \emph{IEEE
  Communications Surveys \& Tutorials}, vol.~19, no.~1, pp. 465--481, 2016.

\bibitem{lokhov2014inferring}
A.~Y. Lokhov, M.~M{\'e}zard, H.~Ohta, and L.~Zdeborov{\'a}, ``Inferring the
  origin of an epidemic with a dynamic message-passing algorithm,''
  \emph{Physical Review E}, vol.~90, no.~1, p. 012801, 2014.

\bibitem{shah2012rumor}
D.~Shah and T.~Zaman, ``Rumor centrality: a universal source detector,'' in
  \emph{Proceedings of the 12th ACM SIGMETRICS/PERFORMANCE joint international
  conference on Measurement and Modeling of Computer Systems}, 2012, pp.
  199--210.

\bibitem{dong2013rooting}
W.~Dong, W.~Zhang, and C.~W. Tan, ``Rooting out the rumor culprit from
  suspects,'' in \emph{2013 IEEE International Symposium on Information
  Theory}.\hskip 1em plus 0.5em minus 0.4em\relax IEEE, 2013, pp. 2671--2675.

\bibitem{chai2021information}
Y.~Chai, Y.~Wang, and L.~Zhu, ``Information sources estimation in time-varying
  networks,'' \emph{IEEE Transactions on Information Forensics and Security},
  vol.~16, pp. 2621--2636, 2021.

\bibitem{watts1998collective}
D.~J. Watts and S.~H. Strogatz, ``Collective dynamics of `small-world'
  networks,'' \emph{Nature}, vol. 393, no. 6684, pp. 440--442, 1998.

\bibitem{newman2018networks}
M.~Newman, \emph{Networks}.\hskip 1em plus 0.5em minus 0.4em\relax Oxford
  university press, 2018.

\bibitem{morone2015influence}
F.~Morone and H.~A. Makse, ``Influence maximization in complex networks through
  optimal percolation,'' \emph{Nature}, vol. 524, no. 7563, pp. 65--68, 2015.

\bibitem{cohen2003efficient}
R.~Cohen, S.~Havlin, and D.~Ben-Avraham, ``Efficient immunization strategies
  for computer networks and populations,'' \emph{Physical Review Letters},
  vol.~91, no.~24, p. 247901, 2003.

\bibitem{liu2016local}
Y.~Liu, Y.~Deng, and B.~Wei, ``Local immunization strategy based on the scores
  of nodes,'' \emph{Chaos: An Interdisciplinary Journal of Nonlinear Science},
  vol.~26, no.~1, p. 013106, 2016.

\bibitem{ren2019generalized}
X.-L. Ren, N.~Gleinig, D.~Helbing, and N.~Antulov-Fantulin, ``Generalized
  network dismantling,'' \emph{Proceedings of the national academy of
  sciences}, vol. 116, no.~14, pp. 6554--6559, 2019.

\bibitem{fan2020finding}
C.~Fan, L.~Zeng, Y.~Sun, and Y.-Y. Liu, ``Finding key players in complex
  networks through deep reinforcement learning,'' \emph{Nature Machine
  Intelligence}, pp. 1--8, 2020.

\bibitem{mugisha2016identifying}
S.~Mugisha and H.-J. Zhou, ``Identifying optimal targets of network attack by
  belief propagation,'' \emph{Physical Review E}, vol.~94, no.~1, p. 012305,
  2016.

\bibitem{braunstein2016network}
A.~Braunstein, L.~Dall’Asta, G.~Semerjian, and L.~Zdeborov{\'a}, ``Network
  dismantling,'' \emph{Proceedings of the National Academy of Sciences}, vol.
  113, no.~44, pp. 12\,368--12\,373, 2016.

\bibitem{stauffer2018introduction}
D.~Stauffer and A.~Aharony, \emph{Introduction to percolation theory}.\hskip
  1em plus 0.5em minus 0.4em\relax CRC press, 2018.

\bibitem{keeling2011modeling}
M.~J. Keeling and P.~Rohani, \emph{Modeling infectious diseases in humans and
  animals}.\hskip 1em plus 0.5em minus 0.4em\relax Princeton university press,
  2011.

\bibitem{barabasi1999emergence}
A.-L. Barab{\'a}si and R.~Albert, ``Emergence of scaling in random networks,''
  \emph{Science}, vol. 286, no. 5439, pp. 509--512, 1999.

\bibitem{molloy1995critical}
M.~Molloy and B.~Reed, ``A critical point for random graphs with a given degree
  sequence,'' \emph{Random structures \& algorithms}, vol.~6, no. 2-3, pp.
  161--180, 1995.

\bibitem{cohen2000resilience}
R.~Cohen, K.~Erez, D.~Ben-Avraham, and S.~Havlin, ``Resilience of the internet
  to random breakdowns,'' \emph{Physical Review Letters}, vol.~85, no.~21, p.
  4626, 2000.

\bibitem{barabasi2016network}
A.-L. Barab{\'a}si \emph{et~al.}, \emph{Network science}.\hskip 1em plus 0.5em
  minus 0.4em\relax Cambridge university press, 2016.

\bibitem{erdds1959random}
P.~Erd{\H{o}}s and A.~R{\'e}nyi, ``On random graphs {I}.'' \emph{Publicationes
  Mathematicae (Debrecen)}, vol.~6, pp. 290--297, 1959.

\bibitem{albert2000error}
R.~Albert, H.~Jeong, and A.-L. Barab{\'a}si, ``Error and attack tolerance of
  complex networks,'' \emph{Nature}, vol. 406, no. 6794, pp. 378--382, 2000.

\bibitem{cohen2001breakdown}
R.~Cohen, K.~Erez, D.~Ben-Avraham, and S.~Havlin, ``Breakdown of the internet
  under intentional attack,'' \emph{Physical Review Letters}, vol.~86, no.~16,
  p. 3682, 2001.

\bibitem{clusella2016immunization}
P.~Clusella, P.~Grassberger, F.~J. P{\'e}rez-Reche, and A.~Politi,
  ``Immunization and targeted destruction of networks using explosive
  percolation,'' \emph{Physical Review Letters}, vol. 117, no.~20, p. 208301,
  2016.

\bibitem{liu2018optimization}
Y.~Liu, X.~Wang, and J.~Kurths, ``Optimization of targeted node set in complex
  networks under percolation and selection,'' \emph{Physical Review E},
  vol.~98, no.~1, p. 012313, 2018.

\bibitem{liu2019framework}
------, ``Framework of evolutionary algorithm for investigation of influential
  nodes in complex networks,'' \emph{IEEE Transactions on Evolutionary
  Computation}, vol.~23, no.~6, pp. 1049--1063, 2019.

\bibitem{fan2019dismantle}
C.~Fan, Y.~Sun, Z.~Li, Y.-Y. Liu, M.~Chen, and Z.~Liu, ``Dismantle large
  networks through deep reinforcement learning,'' in \emph{ICLR representation
  learning on graphs and manifolds workshop}, 2019.

\bibitem{borgs2014maximizing}
C.~Borgs, M.~Brautbar, J.~Chayes, and B.~Lucier, ``Maximizing social influence
  in nearly optimal time,'' in \emph{Proceedings of the twenty-fifth annual
  ACM-SIAM symposium on Discrete algorithms}.\hskip 1em plus 0.5em minus
  0.4em\relax SIAM, 2014, pp. 946--957.

\bibitem{goldenberg2001talk}
J.~Goldenberg, B.~Libai, and E.~Muller, ``Talk of the network: A complex
  systems look at the underlying process of word-of-mouth,'' \emph{Marketing
  letters}, vol.~12, no.~3, pp. 211--223, 2001.

\bibitem{kempe2003maximizing}
D.~Kempe, J.~Kleinberg, and {\'E}.~Tardos, ``Maximizing the spread of influence
  through a social network,'' in \emph{Proceedings of the ninth ACM SIGKDD
  international conference on Knowledge discovery and data mining}, 2003, pp.
  137--146.

\bibitem{cho2011friendship}
E.~Cho, S.~A. Myers, and J.~Leskovec, ``Friendship and mobility: user movement
  in location-based social networks,'' in \emph{Proceedings of the 17th ACM
  SIGKDD international conference on Knowledge discovery and data
  mining}.\hskip 1em plus 0.5em minus 0.4em\relax ACM, 2011, pp. 1082--1090.

\bibitem{leskovec2009community}
J.~Leskovec, K.~J. Lang, A.~Dasgupta, and M.~W. Mahoney, ``Community structure
  in large networks: Natural cluster sizes and the absence of large
  well-defined clusters,'' \emph{Internet Mathematics}, vol.~6, no.~1, pp.
  29--123, 2009.

\bibitem{keeling2003modelling}
M.~Keeling, M.~Woolhouse, R.~May, G.~Davies, and B.~T. Grenfell, ``Modelling
  vaccination strategies against foot-and-mouth disease,'' \emph{Nature}, vol.
  421, no. 6919, pp. 136--142, 2003.

\end{thebibliography}

\end{document}